\pdfoutput=1
\documentclass[aps,pre,twocolumn,superscriptaddress,showpacs]{revtex4}

 \usepackage{graphicx}    
 \usepackage{amsmath}
 \usepackage{amssymb}
\usepackage{multirow}
\usepackage{array}
\usepackage{rotating}

\newcommand{\e}[0]{\epsilon}
\newcommand{\tk}[0]{\tilde{\kappa}}

\newcommand{\tD}[0]{\tilde{\Delta}}
\newcommand{\dd}{\mathrm{d}}
\newcommand{\lm}{l_{\mathrm{m}}}
\newcommand{\ylm}{Y_l^m}
\newcommand{\Eqref}[1]{\eqref{#1}}
\newcommand{\xith}{\xi_{lm}^{(\mathrm{th})}}
\newcommand{\xia}{\xi_{lm}^{(\mathrm{a})}}
\newcommand{\xiae}{\xi_{l'm'}^{(\mathrm{a})*}}
\newcommand{\A}{\chi_{\mathrm{a}}}
\newcommand{\ta}{\tau_\mathrm{a}}
\newcommand{\kt}{\tilde{\kappa}}
\newcommand{\At}{\tilde{\chi}_{\mathrm{a}}}
\newcommand{\taut}{\tilde{\tau}}
\newcommand{\Dt}{\tilde{\Delta}}
\newcommand{\D}{\Delta}
\newcommand{\sth}{\sigma_\mathrm{th}}
\newcommand{\sa}{\sigma_\mathrm{a}}
\newcommand{\Dulm}{\delta \langle \left| u_{lm}\right|^2 \rangle}

\newcommand{\tm}{t_\mathrm{m}^{(l)}}
\newcommand{\tml}[1]{t_\mathrm{m}^{(#1)}}

\newcommand{\lc}{ {l_{\rm c}} }

\begin{document}


\title{Effective tension and fluctuations in active membranes}


\author{Bastien Loubet}
\affiliation{MEMPHYS - Center for Biomembrane Physics, Department of Physics, Chemistry and Pharmacy, University of Southern Denmark, Campusvej 55, 5230 Odense M, Denmark}

\author{Udo Seifert}
\affiliation{Universit\"at Stuttgart, II. Institut f\"ur Theoretische Physik, Pfaffenwaldring 57 / III, D-70550 Stuttgart, Germany}
 
\author{Michael Andersen Lomholt}
\affiliation{MEMPHYS - Center for Biomembrane Physics, Department of Physics, Chemistry and Pharmacy, University of Southern Denmark, Campusvej 55, 5230 Odense M, Denmark}

\date{\today}

\begin{abstract}
We calculate the fluctuation spectrum of the shape of a lipid vesicle or cell exposed to a non-thermal source of noise. In particular we take into account constraints on the membrane area and the volume of fluid that it encapsulates when obtaining expressions for the dependency of the membrane tension on the noise. We then investigate three possible origins of the non-thermal noise taken from the literature: A direct force, which models an external medium pushing on the membrane. A curvature force, which models a fluctuating spontaneous curvature, and a permeation force coming from an active transport of fluid through the membrane. For the direct force and curvature force cases, we compare our results to existing experiments on active membranes.
\end{abstract}

\pacs{87.16.dj, 87.16.D-}

\maketitle

\section{Introduction}

Lipid membranes are an essential part of any biological cell, for example as the interface between the cell and its surrounding environment. The cell is actively engaged in interactions with its environment which means that the membrane is kept out of thermal equilibrium. One of the interesting properties of the membrane are its shape fluctuations, which can influence for instance membrane adhesion \cite{Evans1986,Lipowsky1991,Zidovska2006,Reister2008,Brown2008}. 
The shape fluctuations must be influenced by the activity since it is for example known that the fluctuations of red blood cells depend on the concentration of ATP \cite{levin1991}. Artificial vesicles are a valuable model for living membranes as well as an interesting physical object of its own \cite{Seifert1997}. The simplest fluid vesicles are in thermal equilibrium. Their fluctuations are purely thermal and characterized by the well known Helfrich effective free energy \cite{Helfrich1973}. It contains the bending rigidity, the Gaussian bending rigidity and the tension. The bending rigidity and the Gaussian bending rigidity are material parameters, which depend mainly on the membrane composition, while the tension can be interpreted as a Lagrange multiplier for a constraint on the membrane area \cite{Seifert1995}. The tension is therefore related to the excess area stored in the fluctuations of the membrane shape.
 Increasing the excess area often leads to a response of decreasing tension and vice versa. 
A number of theoretical and experimental works on active membranes consider the active contribution in the membrane dynamical equations disregarding effects on the interplay between tension and excess area of the membrane \cite{Manneville2001,Mdipole,Gov2004a,Sankararaman2002,Lacoste2005}. The correct physical treatment of the renormalization of the tension is an important problem since, for instance, the membrane tension has been shown to affect membrane endocytosis \cite{Dai1997} or mechanosensitive channels \cite{Alberts,Yoshimura2010,Charras2004} as well as influencing the motility of some cells \cite{Batchelder2011}. In this paper we use a simple model for an active force on the membrane to extend the idea of area conservation to active membranes. To do so we compare the passive and active fluctuation spectrums of the same vesicle and we calculate the tension selfconsistently through a constraint on the excess area. We will consider three different physical scenarios for the force that have been studied in the literature: direct force \cite{Gov2004a,Brannigan2006}, curvature force \cite{Mdipole,Manneville2001,Gov2004a} and a case we will call permeation force \cite{Manneville2001,Prost1996}. For the direct force we will consider the example of adenosine triphosphate (ATP) activated cytoskeleton of red blood cells, \cite{Gov2003,Gov2004b,Gov2005} and references therein. For the curvature force we will comment on experiments with giant unilamelar vesicles having active ions pumps included in their membranes \cite{Manneville2001,Girard2005,FarisPRL}. We will give order of magnitude estimates of the effects for the permeation force case.

The outline of the paper is as follows: In Section \ref{sec:Eq}, we recall the dynamical equations governing the fluctuations and the area conservation for passive vesicles. In Section \ref{sec:Act}, we add an active contribution to the dynamic equations for the vesicle and give the corresponding changes in the general expressions for
the fluctuations and the area conservation equations.
In section \ref{sec:Dir}, we discuss the direct force case;
in Section \ref{sec:Cur}, we introduce the curvature force case, 
and in Section \ref{sec:Per} the permeation force case. Finally, we give our conclusions and outlook in Section \ref{sec:conc}.


\section{Equilibrium fluctuations}
\label{sec:Eq}

We begin by recalling the equations of motion for the passive fluctuations of a quasi spherical vesicle taking into account the area and volume constraints \cite{Seifert1995}. We take the following effective free energy for our membrane
\begin{equation}
\label{freeEnergy}
F = \int \dd A \left( 2 \kappa H^2 + \Sigma \right).
\end{equation}
The first term in this equation is the classical Helfrich energy with $\kappa$ being the bending rigidity, $H$ the local mean curvature and the integration being over the whole area of the membrane. The second term is a Lagrange multiplier for the area: the tension $\Sigma$ will be determined selfconsistently such that the total area is conserved. For the quasi-spherical vesicle we will write the radial coordinate of the surface as
\begin{equation}
R(\theta,\phi,t) = R_0 \left( 1 + u(\theta,\phi,t) \right),
\end{equation}
where $R_0$ is the radius defined through the fixed volume $V$ of the vesicle by $V= 4 \pi R_0^3 /3$, and $\theta$ and $\phi$ are the polar and azimuthal angle. We expand the $u(\theta,\phi,t)$ in spherical harmonics as
\begin{equation}u(\theta,\phi,t) = \sum_{l=0}^{\lm} \sum_{m=-l}^{l} u_{lm}(t) \ylm (\theta,\phi),
\end{equation}
where $\lm$ is a cut off value for $l$.
 The equation of motion for the $u_{lm}(t)$ is the normal component of the force balance equation of the membrane. Detailed derivations have been given elsewhere, \cite{Mdipole} or \cite{Seifert1999} for example. We only state the result here
\begin{equation}
\label{EqMotion}
\frac{\eta}{\Gamma_l}\dot{u}_{lm}(t) = - \frac{\kappa}{R_0^3} E_l(\sth) u_{lm}(t) + \xith(t),
\end{equation}
where
\begin{equation}
\Gamma_l \equiv \frac{l(l+1)}{4 l^3 + 6 l^2 -1},
\end{equation}
\begin{equation}
E_l(\sigma) \equiv (l+2)(l-1)\left[l(l+1) + \sigma \right].
\end{equation}
 The left hand side of Eq. \Eqref{EqMotion} is the friction of the surrounding bulk fluid on the membrane with $\eta$ being the viscosity of the fluid and the dot denoting a time derivative. The first term on the right hand side is the elastic restoring force coming from the Helfrich free energy Eq. \Eqref{freeEnergy} where $\sth \equiv \Sigma R_0^2 / \kappa$ is a dimensionless tension. The last term is the thermal noise giving rise to the thermal fluctuations. The properties of this thermal noise are
\begin{subequations}
\label{thnoise}
\begin{equation}
\left\langle \xith(t) \right\rangle = 0
\end{equation}
\begin{equation}
\left\langle \xith(t) \xi_{l'm'}^{(\mathrm{th})}(t')^* \right\rangle = 2 \eta \frac{k_B T}{R_0^3} \frac{1}{\Gamma_l} \delta_{ll'} \delta_{mm'} \delta(t - t'),
\end{equation}
\end{subequations}
where the star denotes complex conjugation, $k_B$ is the Boltzmann constant, $T$ the temperature of the surrounding fluid, $\delta_{ab}$ the Kronecker delta and $\delta(t)$ the Dirac delta function. Here and throughout the paper the brackets mean that we average over noise realizations. From either these equations or the free energy we can calculate the $u_{lm}$ fluctuations to be
\begin{equation}
\left\langle u_{lm}(t) u_{lm}^*(t)\right\rangle_{\mathrm{th}} = \frac{1}{\tk E_l(\sth)},
\end{equation}
where $\tk \equiv \kappa/(k_B T)$ is a reduced bending rigidity.
We now calculate the tension $\sth$ as in \cite{Seifert1995}. We start by writing down the area and volume of the vesicle as a function of the $u_{lm}$'s
\begin{multline}
A = R_0^2 \left(4 \pi \left(1 + \frac{u_{00}}{\sqrt{4 \pi}}\right)^2 + \right. \\ \left.  \sum_{l=1}^{\lm} \sum_{m=-l}^{l} \left|u_{lm}\right|^2 \left(1 + l(l+1)/2 \right) + O(u_{lm}^3)\right)
\end{multline}
\begin{multline}
V = R_0^3 \left(\frac{4 \pi}{3}\left(1 + \frac{u_{00}}{\sqrt{4 \pi}}\right)^3 \right. \\ \left. + \sum_{l=1}^{\lm} \sum_{m=-l}^{l} \left|u_{lm}\right|^2 + O(u_{lm}^3) \right).
\end{multline}
Having a fixed volume $V=4 \pi R_0^3/3$ fixes the amplitude $u_{00}$ as a function of the other $u_{lm}$:
\begin{equation}
u_{00} = - \sum_{l= 1}^{\lm} \sum_{m=-l}^{l} \frac{\left|u_{lm}\right|^2 }{\sqrt{4 \pi}}.
\end{equation}
The excess area, which we will define as $\Delta \equiv (A - 4 \pi R_0^2)/R_0^2$, is then
\begin{equation}
\label{Delta}
\Delta = \sum_{l= 2}^{\lm} \sum_{m=-l}^{l} \left|u_{lm}\right|^2 \frac{1}{2} (l+2) (l-1).
\end{equation}
Note that we did not include a factor of $4 \pi$ in the definition of the excess area for convenience and that the $l=1$ mode, which corresponds to translation, cancel.
To obtain an equation that gives the tension as a function of the excess area we will consider a thermodynamic limit where we can replace $\left|u_{lm}\right|^2$ by $\left\langle \left|u_{lm}\right|^2 \right\rangle_{\mathrm{th}}$ in Eq. \Eqref{Delta}. It then reads
\begin{equation}
\label{DeltaTh}
2 \kt \Delta = \sum_{l=2}^{\lm} \frac{2l +1}{l(l+1) + \sth}.
\end{equation}
This gives a selfconsistent equation for the passive tension $\sth$ that has been studied in \cite{Seifert1995}.

\section{Fluctuations in the active case}
\label{sec:Act}
 To model the activity we add an active random contribution in the force balance equation Eq. \Eqref{EqMotion}. The equation of motion in the active case is then
\begin{equation}
\frac{\eta}{\Gamma_l}\dot{u}_{lm}(t) = - \frac{\kappa}{R_0^3} E_l(\sa) u_{lm}(t) + \xith(t) + \xia(t),
\label{eqDyn}
\end{equation}
where we introduced a new subscript in the symbol for the tension $\sa$ in the active case, indicating that it can be different from the tension $\sth$ in the passive case, and $\xia(t)$ is the noise coming from the active processes. We will take this noise to have the following properties
\begin{subequations}
\label{actnoise}
\begin{equation}
\left\langle \xia(t)\right\rangle = 0
\end{equation}
\begin{equation}
\left\langle \xia(0) \xiae(t)\right\rangle = \A x_l  \delta_{ll'} \delta_{mm'} \exp\left[- \frac{\left|t\right|}{\ta}\right],
\end{equation}
\end{subequations}
where $\ta$ is a characteristic correlation time of the active process studied. It could be on the order of the average time for a pumping cycle for an ion pump for example. $\A$ gives the strength of the noise. It has the dimension of a force per unit area squared (${\rm N^2m^{-4}}$). $x_l$ is a dimensionless quantity that carries the wavenumber dependency of the noise. By spherical symmetry it can only be a function of $l$ and not $m$.
 From the Langevin equation \Eqref{eqDyn} and the noise properties Eqs. \Eqref{actnoise} and \Eqref{thnoise} the $u_{lm}$ fluctuations can be calculated as
 \begin{equation}
 \label{ActFluct}
\left\langle \left| u_{lm}\right|^2 \right\rangle_{\mathrm{a}} = \frac{1}{\kt E_l(\sa)} \left(1 + \At \frac{x_l}{E_l(\sa)}\frac{ \ta }{ \ta + \tm  }\right),
\end{equation}
where
\begin{equation}
	\At = \A \frac{R_0^6}{k_B T \kappa},
\end{equation}
\begin{equation}
	\tm  \equiv \frac{\eta R_0^3}{\kappa \Gamma_l E_l(\sa)}.
\end{equation}
 $\At$ is a non-dimensional measure for the strength of the noise and $\tm$ is the correlation time for mode $l$ of the shape in thermal equilibrium. The fluctuation spectrum is similar to the passive fluctuation spectrum with the active tension $\sa$ replacing the passive tension $\sth$ and an additional positive contribution proportional to the strength of the noise $\At$. The effective equation of area conservation \Eqref{Delta} (with $\left|u_{lm}\right|^2$ replaced by $\left\langle \left|u_{lm}\right|^2 \right\rangle_{\mathrm{a}}$) is
\begin{equation}
\label{DeltaAct}
2 \tk \Delta = \sum_{l=2}^{\lm} \frac{2 l +1}{l(l+1) + \sa} \left(1 + \At \frac{x_l}{E_l(\sa)}\frac{ \ta }{ \ta + \tm  }\right).
\end{equation}
 Equation \Eqref{DeltaAct} is a central result of this paper. It relates the excess area $\D$ and the tension $\sa$ to the parameters describing the activity: $\At$, $x_l$ and $\ta$. We see that the effect of the activity is twofold: it give a direct contribution to the fluctuation spectrum Eq. \Eqref{ActFluct} and it indirectly renormalises the tension through the area constraint Eq. \Eqref{DeltaAct}.
 
 \subsection{Tension renormalization}
 
 Using Eq. \Eqref{DeltaAct} we can deduce some general properties of the tension in our model.
Comparing the equations for the area conservation in both active case, Eq. \Eqref{DeltaAct}, and passive case, Eq. \Eqref{DeltaTh}, we can see that $\sa \geq \sth$ with $\sa = \sth$ only if $\At = 0$. Because we have fixed the area of the membrane then the additional noise due to the active forces will always lead to a counterbalancing increase in the tension.
Also note that in our model the total tension that appear in the fluctuation spectrum, Eq. \Eqref{ActFluct}, is fixed selfconsistently entirely through Eq. \Eqref{DeltaAct}. This implies that any direct contribution to the tension in the force balance equation \Eqref{eqDyn} has no effect on the fluctuation spectrum, since it will be canceled by a counterbalancing tension contribution to keep the excess area conserved. 
In particular the direct contribution to the tension
from the dipole moment of a distribution of force due to an active protein found in \cite{MdipoleTh} does not affect the fluctuation spectrum in the present model of an incompressible membrane (see \cite{Michael2011} for the case of a slightly elastic membrane). However, the quadrupole contribution found in \cite{MdipoleTh} will act like the curvature force discussed here in Section \ref{sec:Cur} and thereby it indirectly renormalizes the tension through the area constraint.

\label{sec:tension}

We now investigate the relation between the active tension $\sa$, the excess area $\Delta$ and the other parameters of the system using Eq. \Eqref{DeltaAct} in different limiting cases. First we distinguish two limits: long and short active correlation time. For long correlation time, formally $\ta \gg \tml{2}$, Eq. \Eqref{DeltaAct} becomes
\begin{multline}
\label{DeltaActSimpl}
\Dt = \frac{5}{6+ \sa}\left(1 + \At \frac{x_2}{E_2(\sa)}\right) \\+\sum_{l=3}^{\lm} \frac{2l+1}{l(l+1) + \sa} \left(1 + \At \frac{x_l}{E_l(\sa)}\right),
\end{multline}
where $\Dt$ is the reduced excess area
\begin{equation}
 \Dt \equiv 2 \Delta \kt = 2 \Delta \frac{\kappa}{k_B T}.
\end{equation}
For a short correlation time, formally $\ta \ll \tml{\lm}$, we get
\begin{multline}
\label{Dst}
\Dt = \frac{5}{\sa + 6}\left(1 + \At \taut x_2 \frac{6}{55}\right) \\+\sum_{l=3}^{\lm} \frac{2l + 1}{\sa + l(l+1)} \left(1 + \At \taut x_l \Gamma_l\right),
\end{multline}
where $\taut$ reads
\begin{equation}
 \taut \equiv \ta \frac{\kappa}{\eta R_0^3}.
\end{equation}
Note that this case corresponds to a timewise Dirac delta correlated active noise, i.e., a white noise. It can be obtained as $\taut \rightarrow 0$ while $\At \rightarrow \infty$ such that $\At \taut$ is finite. In both cases, Eq. \Eqref{DeltaActSimpl} and Eq. \Eqref{Dst}, we have isolated the $l=2$ term as this term diverges when $\sa \rightarrow -6$. In the last term of both equations we will approximate the sum by an integral while the first term will be kept as it is. Using this approach we can invert Eqs. \Eqref{DeltaActSimpl} and \Eqref{Dst} in order to obtain a relation between $\sa$ and the reduced excess area $\Dt$. Similarly to the treatment of the passive case in \cite{Seifert1995} we also break up our results depending on the value of $\sa$: $\sa \rightarrow -6$, $ 1 \ll \sa \ll \lm^2$ and $ 1 \ll \lm^2 \ll \sa$. After expanding in the appropriate limit of $\sa$ and $\lm$ we keep only the dominating term proportional to $\At$ as well as the dominating term that does not depend directly on $\At$ in Eqs. \Eqref{DeltaActSimpl} and \Eqref{Dst}. This procedure gives the results presented in Table \ref{tab:tension} for each of the three cases of force that we will motivate and consider in the following sections. These results are the essential findings of this paper. They give the behavior of the tension in terms of the active parameters $\At$ and $\ta$ in the different physical cases we consider here. These cases, classified by their dependency of $x_l$ on the wavenumber $l$, are: $x_l \sim 1$ (direct force), $x_l \sim l^2$ (permeation force) and $x_l \sim l^4$ (curvature force). 

\section{Direct force}
\label{sec:Dir}

The case $x_l = 1$ corresponds to a noisy force that is being applied locally on the membrane, i.e., a direct random force contribution in Eq. \Eqref{eqDyn}. This kind of noise have been studied previously, mainly to model an active coupling between the cytoskeleton and the membrane in red blood cells \cite{Gov2005,Lin2006,Gov2007}. In this case it represents the punctual force applied on to the membrane (or more precisely, a release of tension) when ATP is used in the process of cutting a link between an actin filament and the membrane. The amplitude of the active correlation can be estimated as
\begin{equation}
\label{directAmp}
	\A \approx \bar{\rho} \left(\frac{\bar{F}_{0}}{R_0}\right)^2,
\end{equation}
where $\bar{F}_{0}$ is on the order of the force applied on the membrane by one active center and $\bar{\rho}$ is the concentration of active centers. In the cytoskeleton case $\bar{\rho}$ will be related to the average density of links activated by ATP and hence it will depend on the ATP concentration \cite{Gov2005}. The active correlation time $\ta$ should be on the order of the release time of an actin filament. Note that the direct force case requires an external source of momentum for Newton's third law to be obeyed; the case cannot be applied to proteins included in the membrane without external attachment, for example ion pumps.

\begin{table*}
\caption{\label{tab:tension}The different behaviors of the tension obtained for the direct force, curvature force and permeation force in the limit of short and long correlation times for the active noise.}

\newcolumntype{C}{ >$ c <$}
\begin{ruledtabular}
\begin{tabular}{c|c|C|C|C}
 & &  \sa \rightarrow -6 &  1 \ll \sa \ll \lm^2  &  1 \ll \lm^2 \ll \sa  \\ \hline
 \multirow{2}{*}{\begin{sideways} Curvature \end{sideways}} &\raisebox{-4mm}{\begin{sideways} $\ta \gg \tml{2}$ \end{sideways}} & \sa= -6 + \dfrac{5}{2 \Dt}\left( 1 + \sqrt{1 + \dfrac{4}{5} \Dt \At} \right) & \sa=\lm^2 \exp \left[- \dfrac{\tD + {\At}/{4}}{1 + {\At}/{4}} \right] & \sa=\dfrac{\lm^2}{2 \tD} \left(  1 + \sqrt{1 + \dfrac{1}{2}\tD \At}\right)\\
\cline{2-2} & \raisebox{-4mm}{\begin{sideways} ${\ta \ll \tml{\lm}}$ \end{sideways}} & \sa= -6 + \dfrac{5}{\tD} \left( 1 + 4 \At \taut \Gamma_2 \right) & \sa = \lm^2 \exp \left(- \tD + \dfrac{\At \taut}{4} \dfrac{\lm^3}{6} \right) & \sa = \dfrac{\lm^2}{\tD} \left( 1 + \dfrac{\At \taut}{40} \lm^3 \right) \\
 \cline{1-2} \multirow{2}{*}{\begin{sideways} Direct \end{sideways}} &\raisebox{-4mm}{\begin{sideways} $\ta \gg \tml{2}$ \end{sideways}} & \sa= -6 + \dfrac{5}{2 \Dt}\left( 1 + \sqrt{1 + \dfrac{1}{5} \Dt \At} \right) & \tD = \ln \left( \dfrac{\lm^2}{\sa} \right) + \dfrac{\At}{\sa^2} \left(\ln \left( \dfrac{\sa}{4} \right)- \dfrac{3}{8}\right) & \sa = \dfrac{\lm^2}{2 \tD} \left( 1 + \sqrt{1 + 4 \dfrac{\At \tD}{\lm^4} \ln\left(\dfrac{\lm^2}{4}\right) } \right) \\
\cline{2-2} &\raisebox{-4mm}{\begin{sideways} $\ta \ll \tml{\lm}$ \end{sideways}} & \sa = -6 + \dfrac{5}{\tD} \left( 1 + \At \taut  \Gamma_2 \right) & \tD = \ln \left( \dfrac{\lm^2}{\sa} \right) + \dfrac{\pi}{4}\At \taut \dfrac{1}{\sqrt{\sa}} & \sa = \dfrac{\lm^2}{\tD} \left( 1 + \dfrac{\At \taut}{2 \lm} \right) \\ \cline{1-2}
 \multirow{2}{*}{\begin{sideways} Permeation \end{sideways}} &\raisebox{-4mm}{\begin{sideways} $\ta \gg \tml{2}$ \end{sideways}} & \sa = -6 + \dfrac{5}{2 \Dt}\left( 1 + \sqrt{1 +  \dfrac{605}{36} \Dt \At } \right) & \tD = \ln \left( \dfrac{\lm^2}{\sa} \right) + 16 \dfrac{\At}{\sa} & \sa = \dfrac{\lm^2}{2 \tD} \left(  1 + \sqrt{1 + 64 \tD \dfrac{\At}{\lm^2}}\right) \\
\cline{2-2} & \raisebox{-4mm}{\begin{sideways} $\ta \ll \tml{\lm}$ \end{sideways}} & \sa = -6 + \dfrac{5}{\Dt}\left( 1 + \dfrac{55}{6} \At \taut  \right) & \sa = \lm^2 \exp \left[ - \Dt + 8 \At \taut  \lm \right] & \sa = \dfrac{\lm^2}{\tD}\left(  1 + \dfrac{8 }{3} \At \taut \lm \right)  \\
\end{tabular}
\end{ruledtabular}
\end{table*}

\subsection{Tension renormalisation}

\begin{figure}
 \includegraphics{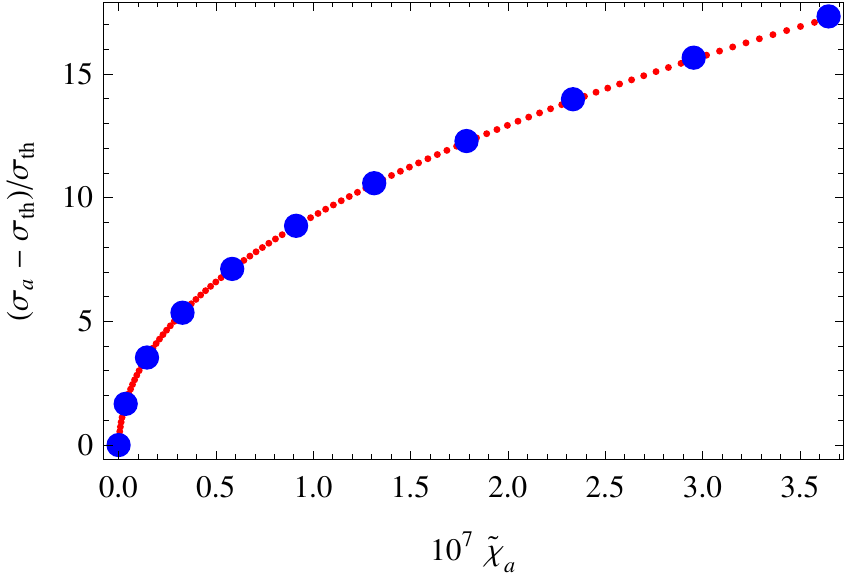}
 \caption{\label{fig:directtension} The relative renormalisation of the tension as a function of the strength of the activity in the direct force case. The horizontal axis represents $\At$ defined by Eq. \Eqref{directAmp} where $\bar{F}_0$ go from $0$ to $0.2\ \mathrm{pN}$. The big circles represent $\sa$'s calculated from Eq. \Eqref{DeltaAct} while the dotted line is $\sa$ calculated from the formula in Table \ref{tab:tension} in the direct force case for $1 \ll \sa \ll \lm^2$ and $\ta \gg \tml{2}$. The paramaters used are: $R_0 \simeq 5\,\mathrm{\mu m}$, $\eta \simeq 9\times 10^{-4}\,\mathrm{kg/m\ s}$, $k_B T \simeq 4\times 10^{-21}\,\mathrm{J}$, $\kappa \simeq 4\, k_B T$, $\ta \simeq 0.1\ \mathrm{s}$, assuming a typical distance between two active centers of $100\ \mathrm{nm}$ we have $\bar{\rho}\simeq 10^{14}\,\mathrm{m^{-2}}$, and taking a membrane thickness of $d \simeq 5\ \mathrm{nm}$ we get $\lm \simeq R_0/d \simeq 1000$. We also chose $\D/4\pi  \simeq 3 \% $ from which we found $\sth \simeq 529$ by numerically solving Eq. \Eqref{DeltaTh}.}
 \end{figure}

The activity will modify the tension through the area constraint as described in Table \ref{tab:tension}.
 We begin our discussion of the application to the cytoskeleton by pointing out that the ATP induced activity, by attempting to increase the membrane fluctuations, increases the membrane tension through the area constraint. Hence the tension increases when the ATP concentration increases. In Fig. \ref{fig:directtension}, we have shown the renormalisation of the tension as a function of the strength of the activity with some typical parameter values for a red blood cell.
 Here we compare this result to another model \cite{Gov2003,Gov2007}. This model introduced a confining potential that act on the membrane in order to model the attachment of the cytoskeleton to the membrane. In addition, the calculation was performed for an almost flat membrane with infinite frame area. In the spherical limit we consider in this paper we can get the flat case limit by keeping only the highest power of $l$ and then replacing $l$ by $ q R_0$ where $q$ is the Fourier transform variable. In this limit $E_l$, including a uniform potential, becomes
\begin{equation}
\frac{\kappa E_l}{R_0^4} \rightarrow \kappa q^4 + \sigma q^2 + \gamma.
\end{equation}
The parameter $\gamma$ is the strength of the confining potential. 
There were no constraints on the area but a effective (increased) tension was considered to come from the confinement of the membrane \cite{Gov2003}. We can give the following two remarks here comparing to our approach. First, in our model a laterally uniform harmonic confining potential over the membrane will decrease the tension compared to a case where there is no confining potential (no attachment to the membrane) because the uniform potential tends to decrease the low wavelength fluctuations; hence the absolute value of the tension will be diminished by the confinement. Secondly, the increase of the tension in our model is due to the effect of the noisy activity through the area constraint. Note that with increasing ATP concentration the strength of the activity $\A$ will increase while the confinement will decrease because the cytoskeleton will have less attachments to the membrane. Thus taking into account the area constraint leads us to conclude that, in our model, when the ATP induced activity increases the tension increases. This contrasts the modelling of \cite{Gov2003,Gov2004b}, where a non-homogenous potential due to the presence of the cytoskeleton induces a positive contribution to the tension, which implies that the tension should decrease with increasing ATP concentration.
 We will show in the next section that our result of an activity induced increase of the tension is not in contradiction with the decrease of the fluctuations observed in ATP depleted red blood cells. Also note that the tension we calculated here is only partially related to the force needed to pull tethers out of cells, as the attachment of the membrane to the cytoskeleton directly contribute to this force as discussed in \cite{Sheetz2001}.

\subsection{Fluctuation renormalisation}

\begin{figure}
 \includegraphics{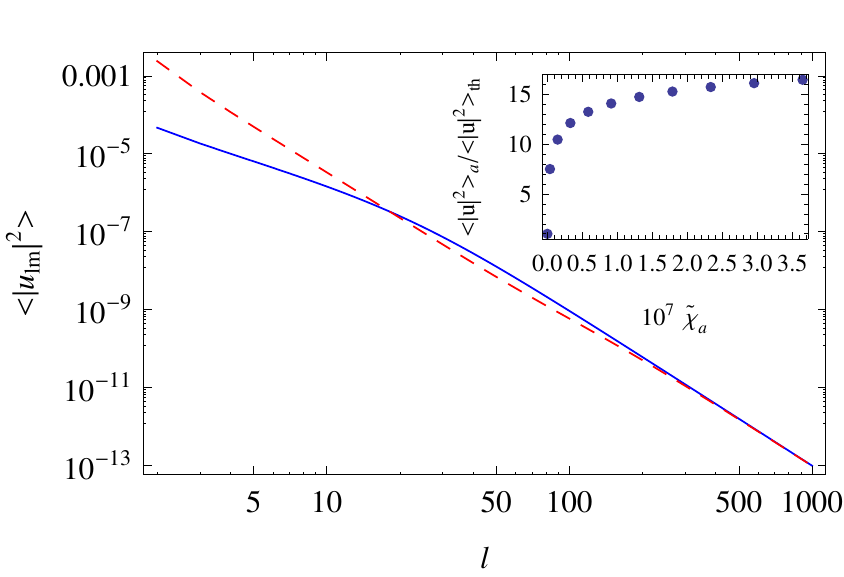}
 \caption{\label{fig:directfluct} Two fluctuation spectrums for the direct force case of the same vesicle with no activity ($\At=0$, continuous line) and with $\bar{F}_0= 0.2\ \mathrm{pN}$ (dashed line). Inset: the relative increase of fluctuations in real space as a function of the active amplitude. The parameters used are the same as in Fig. \ref{fig:directtension}.}
 \end{figure}

In Fig. \ref{fig:directfluct} we compare the passive fluctuation spectrum, $\At =0$, with the corresponding active one for $\bar{F}_0 \simeq 0.2 \,\mathrm{pN}$ and the same parameters as in Fig. \ref{fig:directtension}. We can see that the activity increases the large wavelength (small $l$) fluctuations while decreasing the small wavelength (large $l$) fluctuations; following the general argument laid out in the supplementary material we can show that, in the direct force case for $\sth > 0$, there exists a crossover $l=l_c$ such that the activity always increases the fluctuations for $l< l_c$ and decreases the fluctuations for $l>l_c$. In the inset of Fig. \ref{fig:directfluct} we have plotted the amplitude of the fluctuations in real space $\langle \left| u(\theta,\phi)\right|^2 \rangle$, which can be evaluated as
\begin{equation}
\langle \left| u(\theta,\phi)\right|^2 \rangle=\langle \left| u\right|^2 \rangle = \frac{1}{4 \pi} \sum_{l=0}^{\lm} \sum_{m=-l}^l \langle \left| u_{lm}\right|^2 \rangle.
\end{equation}
This amplitude is independent of $\theta$ and $\phi$ due to rotational symmetry.
From the inset we can see that the real space fluctuation amplitudes can increase by a factor of 4 due to the activity even though the tension is increased and the excess area is conserved. Such an increase of the fluctuations has been observed in \cite{Tuvia1998} by varying the ATP concentration in red blood cells. The effect of the direct force activity is to increase the large wavelength fluctuations (even though the tension is increased) while decreasing the fluctuations in an intermediate range of wavelengths. This increases the observed real space amplitude of the fluctuations while still keeping the total excess area unchanged.

\section{Curvature force}
\label{sec:Cur}

The curvature case has also been studied previously. It is presented as a random active contribution to the spontaneous curvature in \cite{Lin2006}. It can also be seen as a random contribution to the quadrupole moment of a microscopic force density due to active transmembrane proteins \cite{Manneville2001,MdipoleTh}. Note that any second order derivative of a field on the membrane surface, in the force expression in real space, would give a ``curvature force'' noise in our linear theory. Such a noise will lead to the scaling $x_l\sim l^4$ at large $l$. For the subdominant behavior we will follow the behavior found for the quadrupole contribution in \cite{Mdipole} and therefore choose $x_l = (l+2)^2(l-1)^2/4$ for this case and estimate the strength of the noise by
\begin{equation}
\label{Atcurv}
	\A \approx \bar{\rho} \left(\frac{\bar{F}_{2}}{R_0^{3}}\right)^2,
\end{equation}
where $\bar{F}_{2}$ has the unit of force times length squared.
$\bar{\rho}$ is the concentration of active proteins per area of the membrane and $\ta$ is a time scale on the order of the pumping cycle time for an ion pump for example.


\subsection{Micropipette experiments}

In \cite{Manneville2001,Girard2005} vesicles containing active ion pumps have been studied experimentally using a micropipette aspiration technique. In these experiments a vesicle is sucked partially into a micropipette while measuring the change in excess area and the tension applied on the membrane. In both cases they activated the pump externally (by shinning light on them or by injecting ATP). They compared the relation between the excess area and the applied tension for the same vesicle with and without activity. For a passive vesicle the relation between the tension and the excess area is given by Eq. \Eqref{DeltaTh}. In this case a small increment of the excess area ${\rm d}\D$ is related to an increment of the logarithm of the tension ${\rm d}(\ln \sth)$ in the case $1\ll \sth \ll \lm^2$ by
 \begin{equation}
2  \frac{\kappa}{k_B T} {\rm d}\Delta = - {\rm d} \left( \ln  \sth \right).
\end{equation}
For the active membrane they found that this linear relationship between the change in excess area and the logarithm of the tension still holds albeit with a decreased prefactor. This decreased prefactor was assigned to an increased effective temperature $T_\mathrm{eff}$. Using the equations in Table \ref{tab:tension} we can relate the excess area and the tension in the curvature case for $\ta \gg \tml{2}$ and $1 \ll \sa \ll \lm^2$ as
\begin{equation}
	2  \frac{\kappa}{k_B T} {\rm d}\Delta = -\left( 1 + \frac{\At}{4} \right) {\rm d} \left( \ln  \sa \right).\label{eq:actlinear}
\end{equation}
Thus in our model an effective temperature can be defined as $T_\mathrm{eff}/T = 1 + \At/4$.
The difference between our model presented here and the one proposed in \cite{Manneville2001} is that we look at a noisy force distribution with zero mean quadrupole moment while \cite{Manneville2001} had a non-noisy quadrupole force for each protein, which influences the fluctuations through a coupling to the protein density and the curvature of the membrane. Both approaches lead to the linear relationship of Eq. \Eqref{eq:actlinear}, with the correspondance that $\bar{F}_{2} = 2 w \mathcal{P}_a$ where $w$ is defined as the membrane thickness in \cite{Manneville2001} and $\mathcal{P}_a$ is labeled the force dipole.
In the case of bacteriorhodopsin they found experimentally that $T_\mathrm{eff}/T \simeq 2$. By taking $\kappa \simeq 10 k_B T$ and $\bar{\rho} \simeq 10^{16} \,\mathrm{m^{-2}}$ we find from this the quadrupole moment $\bar{F}_2 \simeq 2.6\times 10^{-28}\,\mathrm{J \, m}$. 
It should be noted that the effective temperature $T_{\mathrm{eff}}$ defined here for the curvature case is related to the increase of the fluctuations at small wavelength of the fluctuation spectrum. Indeed taking the limit of the wavenumber $l \rightarrow \infty$ in Eq. \Eqref{ActFluct} we get
  \begin{equation}
\langle \left| u_{lm}\right|^2 \rangle_\mathrm{a} \rightarrow \frac{k_B T}{\kappa l^4} (1+ \frac{\At}{4}) = \frac{k_B T_\mathrm{eff}}{\kappa l^4},
\end{equation}
showing that the same effective temperature appears in the small wavelength region of the fluctuation spectrum. 

\subsection{Videomicroscopy experiment of fluctuating vesicle}

\begin{figure}
 \includegraphics{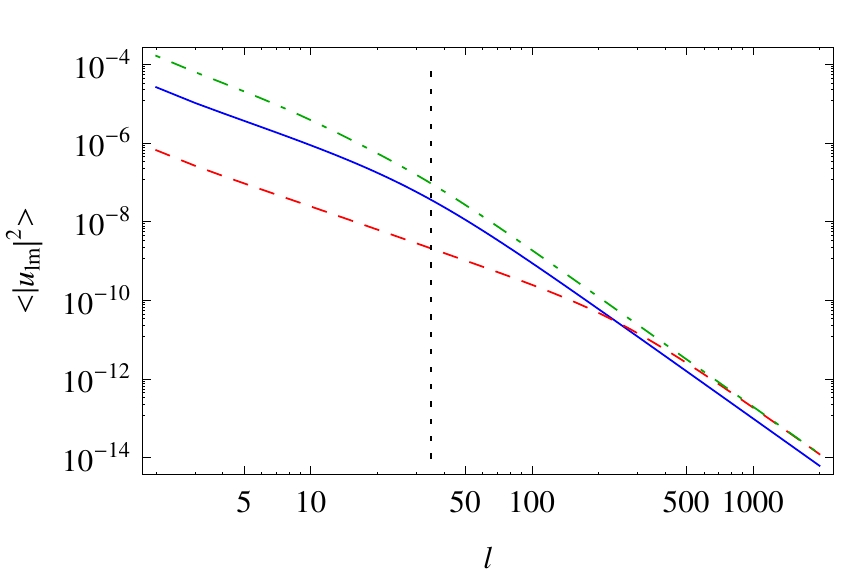}
 \caption{\label{fig:curv} Three fluctuation spectrums for the curvature case. The full line corresponds to the passive case, $\At = 0$, taking the passive tension obtained in \cite{FarisPRL}: $\Sigma_\mathrm{th} \simeq 4\times 10^{-7} \,\mathrm{N\ m^{-1}}$ leading to $\sth \simeq 940$. The dashed line corresponds to the active case with $\At =4$ and a tension $\sa = 3.8\times 10^4$ (giving a conserved  excess area). The dot dashed line is the fluctuation spectrum using values mimicking the experiments, $\At =4$ and $\sa \simeq 130$, see the discussion in the text. The vertical dotted line denotes the approximate maximum wavevector observed in \cite{FarisPRL}. We used the parameters: $\kappa \simeq 10\,k_B T$, $\bar{\rho} \simeq 10^{16} \,\mathrm{m^{-2}}$, $R_0 \simeq 10\,\mathrm{\mu m}$ and $\lm = R_0/d \simeq 2000$.}
 \end{figure}

Another experiment on artificial active membranes was performed in \cite{FarisPRL}. Here both the active and passive fluctuation spectrum of a giant unilamellar vesicle containing the proton pump bacteriorhodopsin was recorded by videomicroscopy, without the vesicle tension being constrained externally. Before discussing their results we will lay out what we would expect in terms of the present theory:
 The short wavelength effective temperature found in the micropipette experiment should still be present in this free vesicle experiment. Thus the active contribution should still increase the short wavelength (large $l$) fluctuations compared to the passive ones. For the membrane to conserve its excess area it needs to compensate for the increase of the small wavelength fluctuations through an increase of the tension such that the large wavelength fluctuations are decreased. This can be seen in Fig. \ref{fig:curv} where  the expected active spectrum (dashed line) is compared to the passive one (full line). In fact for the curvature case it can be proven that the active fluctuation spectrum is always larger than the passive one for small $l$ and then become smaller at values of $l$ above a critical $l=l_c$ (see the supplementary material for a calculation of the $l_c$'s for small $\At$).

If we take the limit of long correlation time: $\ta \gg \tml{2}$ of Eq. \Eqref{ActFluct} for the curvature force case we get
 \begin{equation}
 \label{ActFluctexp}
\left\langle \left| u_{lm}\right|^2 \right\rangle_{\mathrm{a}} = \frac{k_B T}{\kappa E_l(\sa)} \left(1 + \frac{\At}{4} \frac{(l+2)^2(l-1)^2}{E_l(\sa)}\right).
\end{equation}
The flat limit of this equation, i.e., ignoring subdominant terms in $l$ e.g. $(l+2)(l-1)\sim l^2$, is the equation used to fit
the experimentally obtained fluctuation spectrum in \cite{FarisPRL}
if we evaluate $\At$ using Eq. \Eqref{Atcurv}. This is true even though we have considered a shot noise with zero mean while they have considered a mean contribution coupled to the proteins density difference (similarly to the discussion in the last subsection). Or in other words: a very long correlation time for the noise is equivalent to a constant mean contribution in this case. Experimentally they found a significant increase of the fluctuation spectrum at low $l$ values when the proteins were activated, which they interpreted as an active reduction of the tension from $\Sigma_\mathrm{th} \simeq 4\times 10^{-7} \,\mathrm{N\,m^{-1}}$ in the passive case to $\Sigma_\mathrm{a} \simeq 5.3\times 10^{-8} \,\mathrm{N\,m^{-1}}$ in the active case. This gives $\sth = 940$ and $\sa= 130$ for our unitless tensions. The corresponding active fluctuation spectrum calculated from Eq. \Eqref{ActFluctexp}, with the value $\At = 4$ taken from the fit to the micropipette experiment and $\sa= 130$, is plotted in Fig. \ref{fig:curv} as a dot dashed line.
 
 \subsection{Discussion}
 
 The lowering of the tension is necessary to explain the increase of the large wavelength fluctuations observed in \cite{FarisPRL} and the increase in effective temperature is necessary to explain the micropipette experiment \cite{Manneville2001}. But this gives an active fluctuation spectrum consistently larger than the passive one for all wavenumbers.
 These results suggest that the excess area is not conserved in these experiments. We can suggest two different mechanisms in order to explain this increase:
The first explanation would be that the increase of the excess area is due to the elastic properties of the lipid bilayer itself and a very large negative contribution to the tension in the active case.
In this case it is the lipidic part of the membranes area which is expanded due to the additional negative active contribution to the tension.
However, the active contribution to the tension need to be very large in magnitude, at least on the order of $10^{-3}\,{\rm Nm^{-1}}$, to give a visible effect on the fluctuation spectrum due to the high compressional modules of a lipid membrane, see \cite{Michael2011}.

Another explanation could be, for these experiments using bacteriorhodopsin, that the area the proteins take up in the membrane increases when they are active (at least on average in time), since this protein is known to periodically expand during its active cycle \cite{Voitchovsky2009}. In this case the area per lipid head in the lipid region of the membrane would be unchanged and the increase of the excess area would be due to the expansion of the proteins. This would allow the membrane tension to decrease.
 
Assuming $\At \simeq 4$ we can evaluate the excess area $\D_{\rm a}$ required to obtain $\sa=130$. Using $\sth = 940$, $l_m=2000$ and $\kappa= 10\,k_B T$ we find for the passive case $\D_{\rm th}/(4 \pi)\simeq 0.03$. 
From the result for the tension in Table \ref{tab:tension} for the curvature case with $\ta \gg \tml{2}$ and $1 \ll \sa \ll \lm^2$ we obtain
\begin{equation}
\Dt_\mathrm{a} = \left( 1 + \frac{\At}{4}\right) \left(\log\left[\frac{\sth}{\sa}\right] + \Dt_\mathrm{th} \right) - \frac{\At}{4}.
\end{equation}  
With $\sa=130$ and $\At \simeq 4$ we find $\D_{\rm a}/(4 \pi)\simeq 0.07$. One can evaluate the corresponding increase of the area per proteins, $a_i$, between the passive and active case. For a protein concentration of $\bar{\rho} \simeq 10^{16} \,\mathrm{m^{-2}}$ we find $a_i \simeq 4\times 10^{-18}\, \mathrm{m^2}$. This correspond to a $\sim10\% $ increase of the active radius, if we take a protein to occupy a disk of radius $r_p\simeq 2.5\, \mathrm{nm}$ in the passive case.

\section{Permeation force}
\label{sec:Per}

 \begin{figure}
 \includegraphics{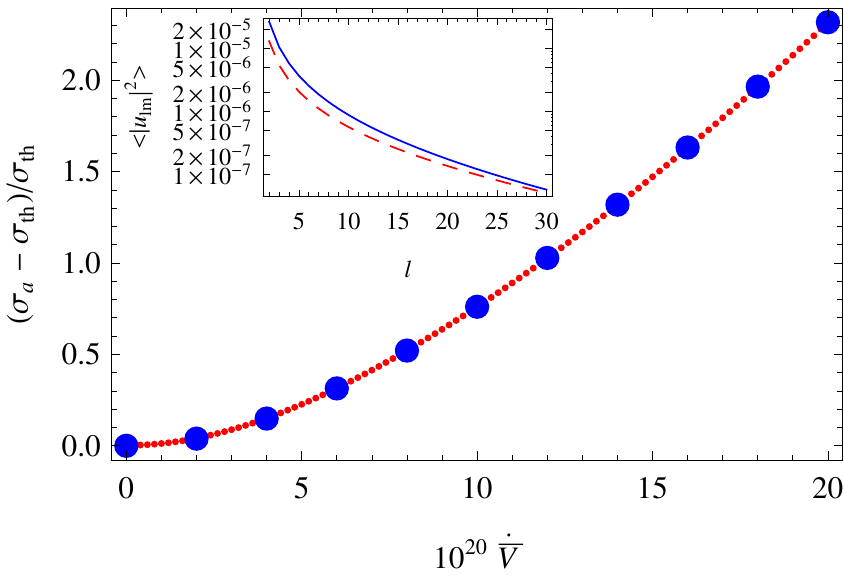}
 \caption{\label{fig:permea} The relative increase in tension as a function of the mean volume transfered through the membrane per unit time for the permeation case. The dotted line is calculated using the equation for $\ta \gg \tml{2}$ in Table \ref{tab:tension} while the big circles are numerical calculations directly from Eq. \Eqref{DeltaAct}. Inset: the first 30 modes of the fluctuation spectrum. The blue line is for the passive case while the red dashed line is for $\dot{\bar{V}}_\mathrm{a} \simeq 1.5 \times 10^{-19}\,\mathrm{m^3/s}$. We used the parameters $\kappa \simeq 10\,k_B T$, $\bar{\rho} \simeq 10^{16} \,\mathrm{m^{-2}}$, $R_0 \simeq 10\,\mathrm{\mu m}$, $\lm = R_0/d \simeq 2000$, $\eta \simeq 10^{-3}\,\mathrm{kg/m\ s}$ and $\sth \simeq 940$.}
 \end{figure}

We will show here that a contribution of the form $x_l \sim l^2$ can be justified through a model of the activity with transfer of fluid between the two sides of the membrane. In Eq. \Eqref{eqDyn} we used a no slip boundary condition to obtain the hydrodynamic stresses on the membrane from the surrounding fluid. The radial component of this boundary condition reads
\begin{equation}
	\dot{R}(\theta,\phi,t) - v^{\mathrm{r}}_{\mathrm{fluid}}(\theta,\phi,R,t) = 0,
\end{equation}
where $R$ is the radial position of the membrane and $v^{\mathrm{r}}_{\mathrm{fluid}}$ is the radial component of the fluid velocity evaluated at the membrane. To obtain the permeation force case we add an active contribution to the boundary condition. It then reads
\begin{equation}
\label{actbound}
	\dot{R}(\theta,\phi,t) - v^{\mathrm{r}}_{\mathrm{fluid}}(\theta,\phi,R,t) = v^\mathrm{r}_\mathrm{a}(\theta,\phi,t),
\end{equation}
where $v^\mathrm{r}_\mathrm{a}$ is the random active contribution of zero mean and with the following correlation after expanding in spherical harmonics
\begin{equation}
\label{permeacorr}
	\left\langle  v^\mathrm{r}_{\mathrm{a},lm}(0)  v^\mathrm{r}_{\mathrm{a},l'm'}(t)  \right\rangle = \bar{\rho} \left( \dot{\bar{V}}_{\rm a} \right)^2 \frac{\delta_{ll'} \delta_{mm'}}{R_0^2} \exp\left[- \frac{\left|t\right|}{\ta}\right].
\end{equation}
Here $\bar{\rho}$ is the mean density of active centers and $\dot{\bar{V}}_{\rm a}$ is on the order of the volume transferred through the membrane per active center per unit time. We have assumed that the active centers are operating independently in Eq. \Eqref{permeacorr}. Using Eq. \Eqref{actbound} instead of the no slip boundary condition we obtain a dynamical equation of the form of Eq. \Eqref{eqDyn} with $\xia(t)= \eta v^\mathrm{r}_{\mathrm{a},lm}(t)/(\Gamma_l R_0)$. The active parameters $\A$ and $x_l$ then read
\begin{equation}
\label{Apermea}
	\A = \bar{\rho}\left(  \frac{\dot{\bar{V}}_{\rm a} \eta}{R_0^2} \right)^2   \quad \mathrm{and} \quad x_l= \left( \Gamma_l \right) ^{-2},
\end{equation}
which gives $x_l\sim l^2$ at large $l$.

In \cite{Manneville2001} a similarly active permeation term was considered for the force dipole model using Darcy's law of permeation. The membrane was taken to have a permeation coefficient $\lambda_\mathrm{p}$ and a force $F_a$ was pushing fluid through it. Our parameters are related to theirs as $\bar{\rho}  \dot{\bar{V}}_{\rm a}^2 \sim  \rho (\lambda_\mathrm{p} F_\mathrm{a})^2$ where $\rho=\bar{\rho}$ is the average pump density on the membrane.
They neglected this contribution in their effective temperature expression for bacteriorhodopsin because the water permeation coefficient through the lipid bilayer is very low, $\lambda_\mathrm{p} \lesssim 10^{-12}\,\mathrm{m^3/(N\,s)}$; however this case could give a significant effect for pumps if they could provide a high enough flow through the membrane. Using Eq.\Eqref{Apermea} in Eq.\Eqref{ActFluct} we found the same equation for the fluctuation spectrum as \cite{Prost1996} for a zero mean active flow in the flat limit.
Unfortunately we are unaware of any experiment relating fluctuations to fluid pumping through the membrane. For a detailed study of the fluctuation spectrum in this permeation force case we refer to the supplementary material.
 
As an example we will evaluate the flow, $\dot{\bar{V}}_{\rm a}$, required to obtain a significant renormalisation of the tension using the parameters of the experiment described in Fig. \ref{fig:curv}.
 We have plotted the renormalisation of the tension, in the limit of long correlation time, as a function of the mean volume per second per protein $\dot{\bar{V}}_\mathrm{a}$ in Fig. \ref{fig:permea}. With the chosen protein concentration we start to have an effect for the renormalisation of the tension around $5\times 10^{-20}\,\mathrm{m^3/s}$ which roughly corresponds to $10^6$ water molecule per $\mathrm{ms}$.
 This order of magnitude for the flow can be obtained for the passive channel aquaporin \cite{Yang1997} for instance.
But we believe it is unlikely that active pumps can attain that level of flow through the membrane.
 Note, however, that decreasing the tension will increase the effect of the activity for a fixed $\At$.

\section{Conclusion}
\label{sec:conc}

In this paper we have investigated the fluctuation spectrum of vesicles with noisy active inclusions. We used a simpler model for the activity than previous studies \cite{Manneville2001,Gov2004a,Mdipole}, since our noise had vanishing mean, but instead we took into account the high compressional modulus of a lipid membrane by an area constraint. We investigated the effect of the area constraint on three physical models for the activity. We showed that our model for the direct force case agree well with previous experimental and theoretical work on red blood cells; albeit with  a modified interpretation for the increase of the tension. For the force curvature case we compared our model to the experiments of \cite{FarisPRL,Manneville2001} and we showed that even though our model recover the effective temperature found in the micropipette experiment, the data for the free vesicle containing active bacteriorhodopsin seems at odds with the conservation of the excess area. This suggests that bactheriorhodopsin might increase the excess area of the vesicle when the proteins are active or that the force curvature case does not completely describe those experiments. We also investigated a permeation force case and gave an order of magnitude of the flow required to obtain a significant tension renormalisation for a specific example. We hope that this study will motivate further theoretical and experimental works on active biomembranes.
\appendix
\section{Supplementary material}

In this supplementary material we investigate a large part of the phase space of the possible behavior of the active fluctuation spectrum, compared to the passive one. 

\renewcommand{\theequation}{S\arabic{equation}}

\subsection{General formalism}
We study the behavior of the active fluctuation spectrum ($\At > 0$) compared to the passive one ($\At =0$) for the same vesicle taking into account the area constraint. To do so we study the sign change as a function of $l$ of
\begin{equation}
	\Dulm \equiv \frac{\langle \left| u_{lm}\right|^2 \rangle_\mathrm{a} - \langle \left| u_{lm}\right|^2 \rangle_\mathrm{th}}{\langle \left| u_{lm}\right|^2 \rangle_\mathrm{th}},
\end{equation}
where $\langle \left| u_{lm}\right|^2 \rangle_\mathrm{a}$ is the active fluctuation spectrum calculated in the main text, Eq. (16), and $\langle \left| u_{lm}\right|^2 \rangle_\mathrm{th}$ is the passive fluctuation spectrum $\At = 0$, Eq. (8) in the main text.
After a little algebra $\Dulm$ reads
\begin{multline}
\label{DulmPropTo}
\Dulm =  \frac{1}{l(l+1) + \sa}\left( (\sth - \sa) \phantom{\frac{x_l}{(l+2)(l-1)}} \right.\\ \left. + \At \frac{x_l}{(l+2)(l-1)} \frac{E_l(\sth)}{E_l(\sa)} \frac{\ta}{\ta + \tm} \right).
\end{multline}
The first term in the parenthesis is negative and constant in $l$ while the last one is positive and depends on $l$. The factor in front of the parenthesis is always positive, hence it does not change the sign of $\Dulm$.
 For the area to be conserved there is at least one sign change of $\Dulm$ in the accessible $l$ range: if the active fluctuations are greater than the passive fluctuations at some $l$ value it must be lower at some other $l$ value to have the same excess area and reciprocally. In the rest of this section we lay out our general strategy to investigate these cross-overs. We will then specialize to the physical cases studied in the main paper in the next three sections.
 
 The first thing we can do is to check if the terms in the parenthesis in \Eqref{DulmPropTo} is monotonously increasing or decreasing in $l$ by studying the sign of its derivative given the physical conditions $\sth > -6$, $\sth < \sa$, $2 \leq l \leq \lm$, $\At > 0$, $\kt >0$ and $\ta >0$. If for these conditions the derivative does not change sign over $2 \leq l \leq \lm$ then the function in the parenthesis of \Eqref{DulmPropTo} is monotonous in $l$. If this is the case then we can conclude that $\Dulm$ has only one sign change in the region $2\leq l \leq \lm$ i.e. there is only one crossover between the active and the passive fluctuation  spectrums.
 We can then deduce if the active fluctuations are larger/smaller than the passive fluctuations before and after the crossovers depending on the sign of the derivative.

 If the function in the parenthesis of \Eqref{DulmPropTo} is not monotonous there might be one or more crossovers depending on the parameter values. In order to study these cases we consider two limiting regimes: long and short active correlation times. For long correlation time, $\ta \gg \tml{2}$, \Eqref{DulmPropTo} reads
\begin{multline}
\label{DulmLong}
	\Dulm =  \frac{1}{l(l+1) + \sa }\left( (\sth - \sa) \phantom{\frac{x_l}{(l+2)(l-1)}} \right.\\ \left. + \At \frac{x_l}{(l+2)(l-1)} \frac{E_l(\sth)}{E_l(\sa)} \right),
\end{multline}
while for short correlation time, $\ta \ll \tml{\lm}$, it reads
\begin{multline}
\label{DulmShort}
	\Dulm =  \frac{1}{l(l+1) + \sa}\left( (\sth - \sa) \phantom{\frac{x_l}{(l+2)(l-1)}} \right.\\ \left. + \At \taut \frac{x_l}{(l+2)(l-1)} E_l(\sth) \Gamma_l \right).
\end{multline}
Then we can again try to differentiate the terms in parenthesis of \Eqref{DulmLong} and \Eqref{DulmShort} to check if they are monotonous. If this is the case then the above reasoning apply (there is only one $l_c$ etc...).

If the function is not monotonous then we resort to an expansion in small $\At$, the precise meaning of which depends on the case studied. We expand the active tension $\sa$ as
\begin{equation}
	\sa = \sth + \e,
\end{equation}
where $\e \ll \sth$ is the difference between the active and passive tension. $\e$ is related to $\At$ through the excess area constraint. To first order in both $\e$ and $\At$ we can rewrite Eq. (19) in the main text, in the limit $\ta \gg \tml{2}$, as
\begin{multline}
\label{firstorder1}
	0 = - \e \sum_{l=2}^{\lm} \frac{2 l + 1}{\left[ l(l+1)+ \sth \right]^2} \\+ \At \sum_{l=2}^{\lm} \frac{2 l+1  }{l(l+1) + \sth} \frac{x_l}{E_l(\sth)},
\end{multline}
and in the limit $\ta \ll \tml{\lm}$ we get
\begin{multline}
\label{firstorder2}
	0 = - \e \sum_{l=2}^{\lm} \frac{2 l + 1}{\left[l(l+1)+ \sth \right]^2} \\ + \At \taut \sum_{l=2}^{\lm} \frac{2 l+1 }{l(l+1) + \sth } x_l \Gamma_l .
\end{multline}
We have assumed that the passive tension $\sth$ satisfies Eq. (13) in the main text with the same excess area as in the active case. $\At=0$ implies $\e=0$. The relation between the two then reads
\begin{equation}
\label{defc1}
	\e = c_1 \At,
\end{equation}
where $c_1$ is the first order coefficient in an expansion of $\e$ in $\At \ll \sth / c_1$. Using the Mac Laurin formula we can calculate the sum in \Eqref{firstorder1} and \Eqref{firstorder2}. We then consider three different cases: $\sth \rightarrow -6$, $1 \ll \sth \ll \lm^2$ and $1 \ll \lm^2 \ll \sth$ and we expand $c_1$ in the appropriate power of $\sth$ and $\lm$. The lowest relevant order of $c_1$ are shown in Table \ref{tab:coef}.

Next we search for the $l= \lc$'s where the magnitude of the active fluctuations cross the passive ones. To do so we solve the equation $\Dulm = 0$ to first order in $\e$ using $\sa = \sth + c_1 \At$. For $\ta \gg \tml{2}$  the $\lc$ are solutions of
\begin{equation}
\label{lchigh}
	c_1 = \frac{x_l}{(l+2)(l-1)},
\end{equation}
while for $\ta \ll \tml{\lm}$ we have
\begin{equation}
\label{lclow}
	c_1 = x_l \Gamma_l \left[ l(l+1) + \sth \right].
\end{equation}
 Note that the crossover values of $\lc$ are independent of $\At$ and $\taut$ for small $\At$. The $\lc$ values we found are shown in Table \ref{tab:coef}. We can get the lowest order of $c_1$ in an expansion in $\sth$ and $\lm$ directly from the equations in Table 1 of the main text. Unfortunately the lowest order is not always sufficient to determine a non trivial $l_c$ value.

In the cases where the term in the parenthesis of $\Dulm$ was not monotonous and $\At > \sth/c_1$ we have calculated the spectrum numerically (in addition to the first order expansion).
 Note that when $\At \rightarrow \infty$ the excess area conservation, Eq. (19) in the main text, tells us that the ratio $\At/\sa^2$ for $\ta \gg \tml{2}$ and $\At/\sa$ for $\ta \ll \tml{\lm}$ is constant. This can be observed also in the last column of Table 1 in the main text. Inserting this result in Eq. (16) in the main text we can see that there is a point, when $\At$ is sufficiently large, where the fluctuation spectrum is not changing significantly anymore as $\At$ increases (note that the tension $\sa$ is still increasing with $\At$). We can then investigate the fluctuation spectrum beginning with a small value of $\At \ll \sth / c_1$ and then increase its value until the fluctuation spectrum do not change significantly anymore.
 
We now investigate each specific case following the method we laid out in this section.

\begin{sidewaystable*}
\caption{ The coefficients $c_1$ (see \Eqref{defc1} in the text) and the crossover wavenumbers $\lc$ for the different cases we investigated.}
\label{tab:coef}
\begin{ruledtabular}
{
\newcolumntype{C}{ >{$}c <{$}}
\begin{tabular}{c|c|C C|C C|C C}
 & &  \multicolumn{2}{|c|}{$\sth \rightarrow -6$} &  \multicolumn{2}{|c|}{$1 \ll \sth \ll \lm^2$}  &  \multicolumn{2}{|c|}{$1 \ll \lm^2 \ll \sth$}  \\ \hline
 & &	c_1		&	\lc		 &		c_1 	& 	\lc		 &		c_1	 &	 \lc		 \\ \hline
 \multirow{2}{*}{\begin{sideways} Curvature \end{sideways}} &\raisebox{-4mm}{\begin{sideways} $\ta \gg \tml{2}$ \end{sideways}} &
	1		&	\gtrsim 2		 & \dfrac{\sth}{4}\left( \ln\left( \dfrac{\lm^2}{\sth}\right)  - 1  \right) & \sqrt{\sth} \sqrt{\ln\left( \dfrac{\lm^2}{\sth} \right) - 1} & \dfrac{\lm^2}{8} & \dfrac{\lm}{\sqrt{2}}    \\
\cline{2-2} &\raisebox{-4mm}{\begin{sideways} $\ta \ll \tml{\lm}$ \end{sideways}} & 
	\dfrac{24}{55} (6 + \sth)+ \dfrac{\lm^3}{120}(6+\sth)^2	&	\gtrsim 2 	 & \dfrac{1}{24} \sth \lm^3 &	\sqrt[5]{\dfrac{2}{3} \sth \lm^3}	& \sth \dfrac{\lm^3}{40}	& \sqrt[3]{\dfrac{2}{5} \lm} \\
 \cline{1-2} \multirow{2}{*}{\begin{sideways} Direct \end{sideways}} &\raisebox{-4mm}{\begin{sideways} $\ta \gg \tml{2}$ \end{sideways}} &
\dfrac{1}{4}	&	\gtrsim 2	&	\dfrac{1}{\sth}\left( \ln\left( \dfrac{\sth}{4} \right) - \dfrac{3}{8}  \right) 	&	\sqrt{\dfrac{\sth}{\ln\left( \sth/4 \right) - 3/8}} & \dfrac{1}{\lm^2} \left( \ln\left( \dfrac{\sth}{4} \right) + \dfrac{5}{8} \right) &	\sqrt{\dfrac{\lm^2}{\ln\left( \sth/4 \right) + 5/8}} \\
\cline{2-2} &\raisebox{-4mm}{\begin{sideways} $\ta \ll \tml{\lm}$ \end{sideways}}
&	\dfrac{6}{55} ( 6 + \sth)	& \gtrsim2	&	\dfrac{\pi}{4} \sqrt{\sth}	&	\dfrac{\pi}{2}\sqrt{\sth}\left( 1 \pm \sqrt{1-\dfrac{4}{\pi^2}} \right) &	\dfrac{\sth}{2 \lm}	&	\dfrac{\lm}{2} \\ \cline{1-2}
 \multirow{2}{*}{\begin{sideways} Permeation \end{sideways}} &\raisebox{-4mm}{\begin{sideways} $\ta \gg \tml{2}$ \end{sideways}} &
\dfrac{3025}{144}	&	\gtrsim 2	&	16 + 20 \dfrac{\ln \sth}{\sth}	&	\sqrt{\dfrac{\sth}{\ln \sth}}	&	16 + 20	\dfrac{\ln \lm^2}{\lm^2}&	\sqrt{\dfrac{\lm^2}{\ln \lm^2}}	 \\
\cline{2-2} & \raisebox{-4mm}{\begin{sideways} $\ta \ll \tml{\lm}$ \end{sideways}} &
\dfrac{55}{6}(6+\sth)+\dfrac{8}{5}\lm (6+\sth)^2	&	\gtrsim 2	&	8\sth\lm	&	\sqrt[3]{2\sth\lm}	&	\dfrac{8}{3}\sth\lm	&	\dfrac{2}{3} \lm \\
\end{tabular}
}
\end{ruledtabular}
\end{sidewaystable*}

\subsection{Curvature force}

In the curvature case the term in parenthesis in \Eqref{DulmPropTo} is always increasing in $l$. $\Dulm$ goes from being negative to being positive as $l$ increases, i.e., the active fluctuation spectrum is smaller than the passive one at low $l$ and larger at larger $l$ with only one crossover between the active and the passive fluctuation spectrum. It can also be seen (less generally) this way: if we look at the active fluctuations, Eq. (16) in the main text, for the curvature case and take the limit $l \rightarrow \infty$, for sufficiently large $\ta$ and assuming $\sa \ll \lm^2$, we get
\begin{equation}
	\langle \left| u_{lm}\right|^2 \rangle_\mathrm{a} \sim \frac{1}{\kt l^2 (l^2 + \sa)} (1+\At \frac{l^2}{4(l^2+\sa)}),
\end{equation}
where we can see that the activity increases the large $l$ fluctuations. The only way the membrane can compensate for this increase is by making the low $l$ fluctuations smaller through an increase in the tension. This is also the case in the limit $\ta \ll \tml{\lm}$ albeit that we get a different $l$ dependency. In the limit $l \rightarrow \infty$  we get 
\begin{equation}
	\langle \left| u_{lm}\right|^2 \rangle_\mathrm{a} \sim \frac{1}{\kt l^2 (l^2 + \sa)} (1+\At \taut \frac{l^3}{16}).
\end{equation}
This increase of the large wavenumber fluctuations can be seen as the activity dragging the excess area toward large $l$.
In the curvature case the behavior of the active fluctuation spectrum relative to the passive one are similar in between short $\ta \ll \tml{\lm}$ and long $\ta \gg \tml{2}$ correlation time. The form of the fluctuation spectrum changes, however, as the $l$ dependency of the active part of the fluctuation spectrum changes in between short and long active correlation time.

\begin{figure*}[h]
\begin{center}
\includegraphics[scale=1]{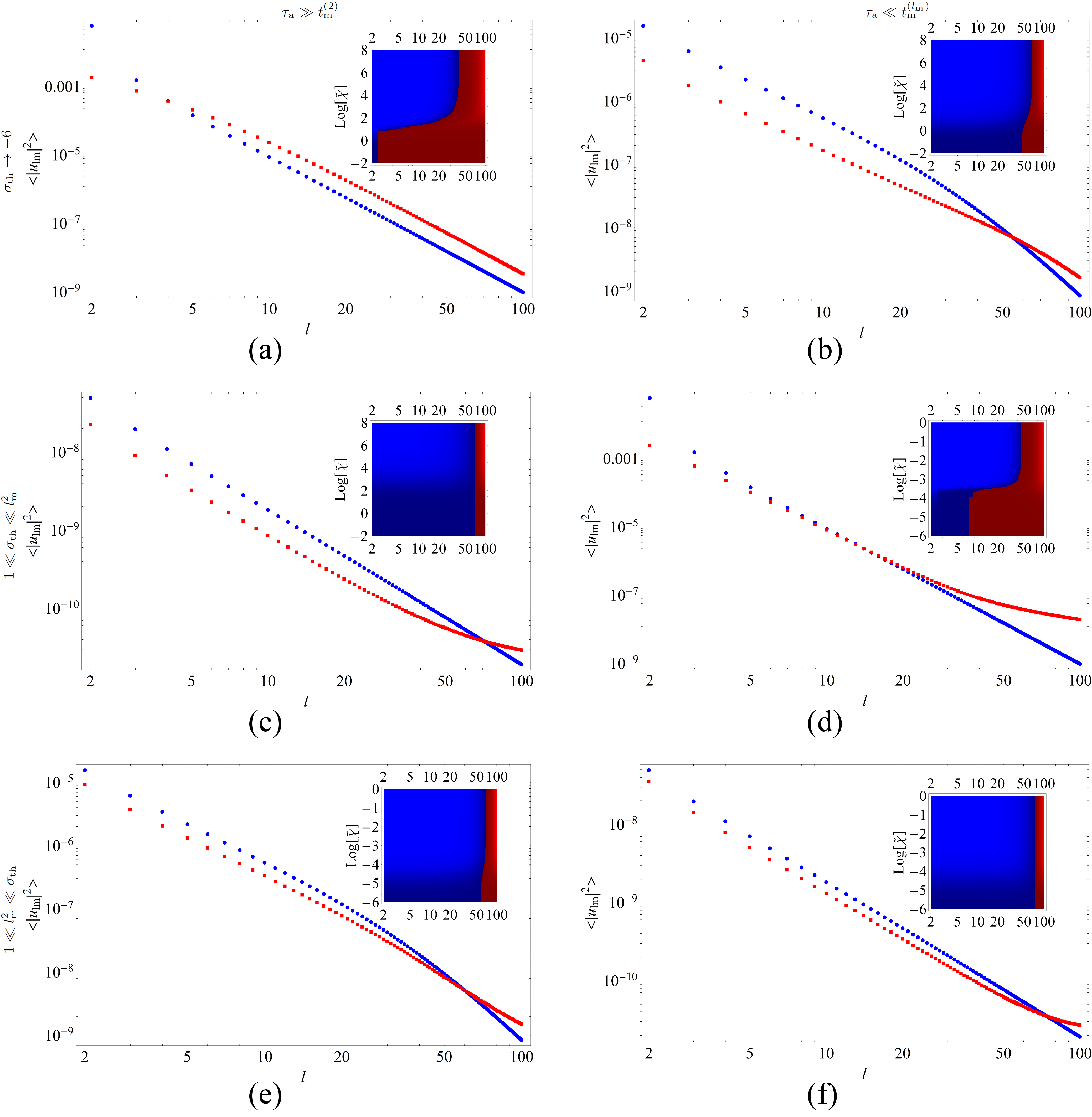}
\caption{\label{fig:curv2} The fluctuation spectrum as a function of $l$ with logarithmic scale on both axis for the curvature force case. The passive fluctuations are in blue and the active fluctuations are in red. For all the subfigures we used $\lm = 100$ and $\kt =10$. Figures (a), (b) and (c) are for $\ta \gg \tm$ while figures (d), (e) and (f) are for $\ta \ll \tm$. Figures (a) and (d) are for $\sth \rightarrow -6$. Here we used $\D = 1$ which gives $\sth \simeq -5.6$, $\At \simeq 10$ for (a) and $\At \simeq0.3\ 10^{-3}$ for (d). Figures (b) and (e) are for $1 \ll \sth \ll \lm^2$. Here we used $\D = 0.1$ which gives $\sth \simeq 1592$, $\At \simeq 10$ for (b) and $\At \simeq1.5\ 10^{-5}$ for (e). Figures (c) and (f) are for $1 \ll \lm^2 \ll \sth$. Here we used $\D = 10^{-3}$ which gives $\sth \simeq 5\  10^5$, $\At \simeq 10^3$ for (c) and $\At \simeq1.5\ 10^{-5}$ for (f). Inset of the figure. is a density plot of the $\Dulm$ with the $\ln l$ along the horizontal axis and the base 10 logarithm of the amplitude of the activity in vertical axis. The graph is red when the active fluctuations are larger than the passive ones and is blue when the passive fluctuations are larger. The more light the color the larger is the absolute value of $\Dulm$. }
\end{center}
\end{figure*}

We show an example in Fig. \ref{fig:curv2} for all the different limits of the correlation time and the tension for purposes of comparing to the direct and permeation force cases.

\subsection{Direct force}
In the direct force case the situation is more complex than in the curvature force case and depend on the sign of both $\sth$ and $\sa$. We will also distinguish between the long and short active correlation time as previously introduced.

\subsubsection{Long correlation time}
\begin{figure*}[h]
\begin{center}
\includegraphics[scale=1]{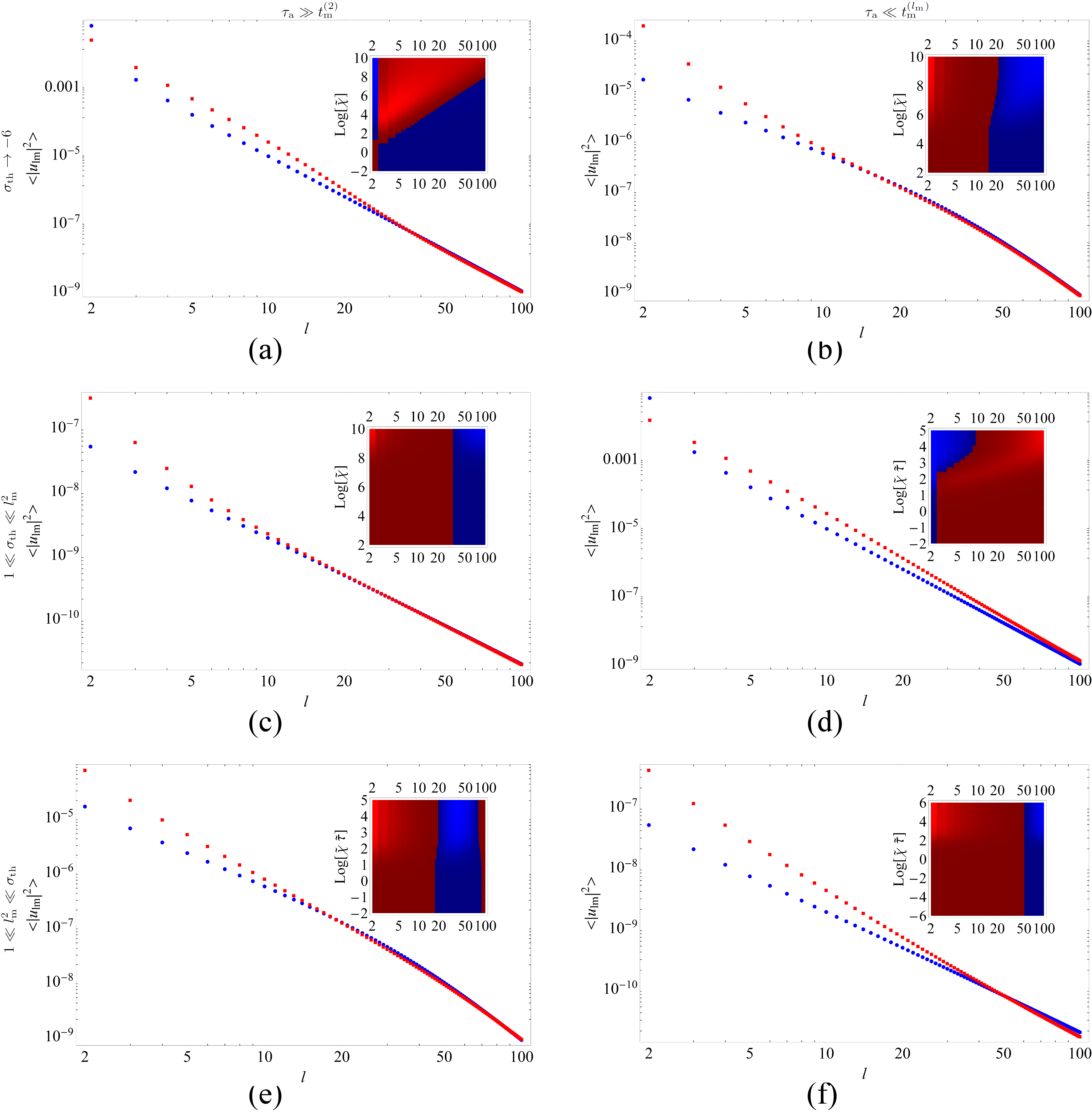}
\caption{\label{fig:direct}The correlation as a function of $l$ in a log log plot for the direct force case. We used the same color coding and parameters as in Fig. 1 except for: (a) $\At \simeq10^6$, (b) $\At \simeq 10^5$, (c) $\At \simeq 10^7$, (d) $\At \simeq 10^2$, (e) $\At \simeq 10^2$ and (f) $\At \simeq 10^2$.}
\end{center}
\end{figure*}

Examples of the fluctuation spectrum for the different limits of the tension have been plotted in (a), (b) and (c) of Fig. \ref{fig:direct}.
 In this case for $\sth \gg 1$ (and hence $\sa>0$) the last term in equation \Eqref{DulmPropTo} is always decreasing in $l$ and we conclude that for $\sth \gg 1$ the active fluctuations are larger than the passive ones at low $l$ and smaller at larger $l$. We stress here that this increase of the fluctuations at small $l$ is present even though the active tension increases. Again we can see this by looking at the $l \rightarrow \infty$ limit
\begin{equation}
	 \langle \left| u_{lm}\right|^2 \rangle_\mathrm{a} \rightarrow \frac{1}{\kt l^2 (l^2 + \sa)} \left( 1+\At \frac{1}{l^2(l^2+\sa)}\right).
\end{equation}
 The part proportional to $\At$ is quickly damped as $l$ increases and the fluctuation spectrum goes under the passive one as $\sa > \sth$ whereas for small $l$ the active noise directly increases the fluctuations.

 In the case $\sth \rightarrow -6$ the small $\At$ expansion tells us that the active fluctuations are larger than the passive ones at $l=2$ and smaller for all larger $l$. In (a) of Fig. \ref{fig:direct} we have put the fluctuation spectrum for a negative $\sth$. In order to interpret this graph we first note that when $\sth \rightarrow -6$ more excess area are stored in the $l=2$ mode. Switching on the activity will then reduce the excess area stored in the $l=2$ mode and increase the area stored in larger $l$. One can say that the activity drag the fluctuations from the $l=2$ mode toward larger $l$ mode.

\subsubsection{Short correlation time}
Examples of the fluctuation spectrum for the different limits of the tension have been plotted in (d), (e) and (f) of Fig. \ref{fig:direct}.
 In this case the only case where the term in the parenthesis of \Eqref{DulmShort} is monotonous is the case where $\sth < 0$. Then $\Dulm$ always goes from negative to positive as $l$ increases. We observe again that the activity drags the excess area toward larger $l$.

In the case $\sth >0$ the small $\At$ expansion gives that for the case $1 \ll \sth \ll \lm^2$ there are two crossover values, both are given in Table \ref{tab:coef}. With the parameters we used the fluctuation spectrum keep the two crossovers at large $\At$ albeit with an increased $l$ value for the crossover, see (e).

For the case $1 \ll \lm^2 \ll \sth$ the second positive root of \Eqref{lclow} becomes larger than $\lm$ and there is only one crossover. The active fluctuations are larger than the passive ones for $l<\lc$ and smaller for $l>\lc$. The existence second root of \Eqref{lclow} is the reason we could not conclude on the behaviors of the fluctuations using the derivative of the term in the parenthesis of \Eqref{DulmShort}. Then the function in the parenthesis of $\Dulm$ is not guaranteed to be monotonous in the whole $l$ range, see (f).

\subsection{Permeation force}

\subsubsection{Long correlation time}

Examples of the fluctuation spectrum for the different limits of the tension have been plotted in (a), (b) and (c) of Fig. \ref{fig:permea2}.
In this case no general conclusion can be derived from the derivative of the term in the parenthesis of \Eqref{DulmLong} so we resort to the expansion in $\At$ for all the cases. All cases are found to have one crossover with active fluctuations larger than the passive ones for $l<\lc$ for small $\At$, but they show different behaviors at large $\At$.

For the case $\sth \rightarrow -6$ and small $\At$ most of the fluctuations are stored in the $l=2$ mode and we observe a similar behavior as in the curvature and direct force cases, i.e., for a very large excess area the activity seems to drag the excess area toward large $l$ values, see (a).

For the case $1 \ll \sth \ll \lm^2$ we can see that a new crossover appears (the root of $\Dulm$ become smaller than $\lm$ ) when $\At$ increases to $\simeq 1$ and the first crossover disappears (the root of $\Dulm$ become smaller than $2$) when $\At$ reach $\simeq 10^2$ resulting in a very different behavior depending on the value of the amplitude of the noise $\At$, see (b). We note that the small $\At$ behavior is similar to the direct force case as the activity directly increases the low wavenumber fluctuations, whereas the large $\At$ behavior is more similar to the curvature force case where the activity increases the large wavenumber fluctuations.

For the case $1 \ll \lm^2 \ll \sth$ we see again a second crossover appearing as $\At$ reach $10^4$ but the first crossover do not disappear. Instead we get a large increase of the $l=2$ amplitude as $\At$ increases further, see (c).
 Note that we can make the second crossover disappear by decreasing the excess area (data not shown).

\subsubsection{Short correlation time}

\begin{figure*}[h]
\begin{center}
\includegraphics[scale=1]{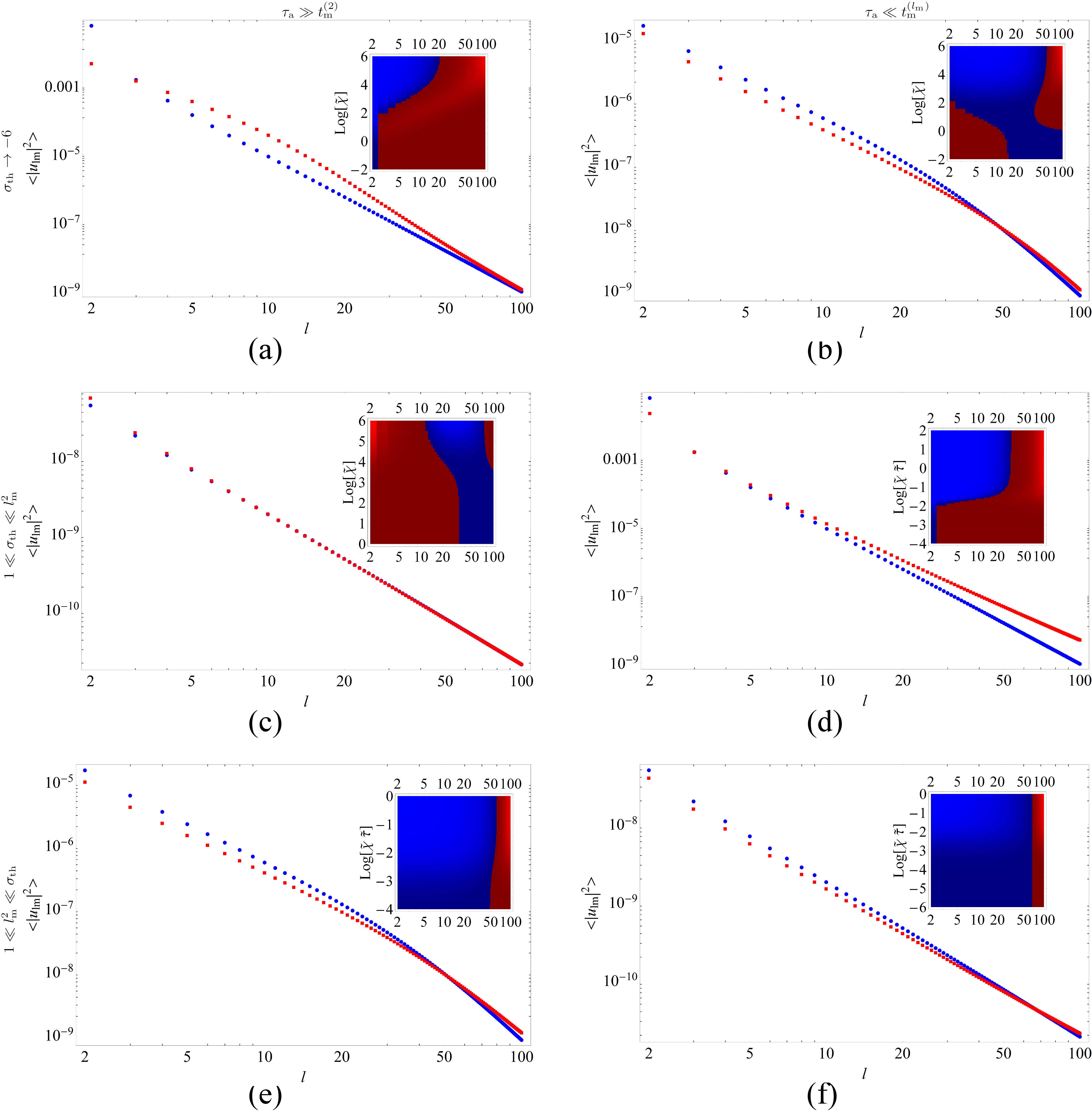}
\caption{\label{fig:permea2}The correlation as a function of $l$ in a log log plot for the permeation force case. We used the same color coding and parameters as in Fig. 1 except for: (a) $\At \simeq10^2$, (b) $\At \simeq 10^3$, (c) $\At \simeq 10^6$, (d) $\At \simeq 10^{-2}$, (e) $\At \simeq 10^{-3}$ and (f) $\At \simeq 10^{-3}$. }
\end{center}
\end{figure*}

Examples of the fluctuation spectrum for the different limits of the tension have been plotted in (d), (e) and (f) of Fig. \ref{fig:permea2}. 
In this case the second term in \Eqref{DulmShort} is always increasing with $l$ and we can conclude in the same way as for the curvature case. The active fluctuations will be lower than the passive fluctuations for low $l<l_c$ and higher for high $l>l_c$. This case is similar to the short correlation time limit of the curvature case for the following reason: if we calculate the $l \rightarrow \infty$ limit of the active fluctuation spectrum, Eq. (16) in the main text, for the permeation force case we get
\begin{equation}
	\langle \left| u_{lm}\right|^2 \rangle_\mathrm{a} \rightarrow \frac{1}{\kt l^2 (l^2 + \sa)} (1+\At \taut 4 l).
\end{equation}
The large wavelength fluctuations are again increased by the activity and the small $l$ fluctuations are correspondingly decreased through the tension to compensate.

\bibliography{biblio}

\begin{thebibliography}{35}
\expandafter\ifx\csname natexlab\endcsname\relax\def\natexlab#1{#1}\fi
\expandafter\ifx\csname bibnamefont\endcsname\relax
  \def\bibnamefont#1{#1}\fi
\expandafter\ifx\csname bibfnamefont\endcsname\relax
  \def\bibfnamefont#1{#1}\fi
\expandafter\ifx\csname citenamefont\endcsname\relax
  \def\citenamefont#1{#1}\fi
\expandafter\ifx\csname url\endcsname\relax
  \def\url#1{\texttt{#1}}\fi
\expandafter\ifx\csname urlprefix\endcsname\relax\def\urlprefix{URL }\fi
\providecommand{\bibinfo}[2]{#2}
\providecommand{\eprint}[2][]{\url{#2}}

\bibitem[{\citenamefont{Evans and Parsegian}(1986)}]{Evans1986}
\bibinfo{author}{\bibfnamefont{E.~A.} \bibnamefont{Evans}} \bibnamefont{and}
  \bibinfo{author}{\bibfnamefont{V.~A.} \bibnamefont{Parsegian}},
  \bibinfo{journal}{Proc. Natl. Acad. Sci.} \textbf{\bibinfo{volume}{83}},
  \bibinfo{pages}{7132} (\bibinfo{year}{1986}).

\bibitem[{\citenamefont{Lipowsky}(1991)}]{Lipowsky1991}
\bibinfo{author}{\bibfnamefont{R.}~\bibnamefont{Lipowsky}},
  \bibinfo{journal}{Nature} \textbf{\bibinfo{volume}{349}},
  \bibinfo{pages}{475} (\bibinfo{year}{1991}), ISSN
  \bibinfo{issn}{{0028-0836}}.

\bibitem[{\citenamefont{Zidovska and Sackmann}(2006)}]{Zidovska2006}
\bibinfo{author}{\bibfnamefont{A.}~\bibnamefont{Zidovska}} \bibnamefont{and}
  \bibinfo{author}{\bibfnamefont{E.}~\bibnamefont{Sackmann}},
  \bibinfo{journal}{Phys. Rev. Lett.} \textbf{\bibinfo{volume}{96}},
  \bibinfo{pages}{048103} (\bibinfo{year}{2006}).

\bibitem[{\citenamefont{Reister-Gottfried
  et~al.}(2008)\citenamefont{Reister-Gottfried, Sengupta, Lorz, Sackmann,
  Seifert, and Smith}}]{Reister2008}
\bibinfo{author}{\bibfnamefont{E.}~\bibnamefont{Reister-Gottfried}},
  \bibinfo{author}{\bibfnamefont{K.}~\bibnamefont{Sengupta}},
  \bibinfo{author}{\bibfnamefont{B.}~\bibnamefont{Lorz}},
  \bibinfo{author}{\bibfnamefont{E.}~\bibnamefont{Sackmann}},
  \bibinfo{author}{\bibfnamefont{U.}~\bibnamefont{Seifert}}, \bibnamefont{and}
  \bibinfo{author}{\bibfnamefont{A.~S.} \bibnamefont{Smith}},
  \bibinfo{journal}{Phys. Rev. Lett.} \textbf{\bibinfo{volume}{101}},
  \bibinfo{pages}{208103} (\bibinfo{year}{2008}).

\bibitem[{\citenamefont{Brown}(2008)}]{Brown2008}
\bibinfo{author}{\bibfnamefont{F.~L.~H.} \bibnamefont{Brown}},
  \bibinfo{journal}{Annu. Rev. Phys. Chem.} \textbf{\bibinfo{volume}{59}},
  \bibinfo{pages}{685} (\bibinfo{year}{2008}).

\bibitem[{\citenamefont{Levin and Korenstein}(1991)}]{levin1991}
\bibinfo{author}{\bibfnamefont{S.}~\bibnamefont{Levin}} \bibnamefont{and}
  \bibinfo{author}{\bibfnamefont{R.}~\bibnamefont{Korenstein}},
  \bibinfo{journal}{Biophys. J.} \textbf{\bibinfo{volume}{60}},
  \bibinfo{pages}{733} (\bibinfo{year}{1991}).

\bibitem[{\citenamefont{Seifert}(1997)}]{Seifert1997}
\bibinfo{author}{\bibfnamefont{U.}~\bibnamefont{Seifert}},
  \bibinfo{journal}{Adv. Phys.} \textbf{\bibinfo{volume}{46}},
  \bibinfo{pages}{13} (\bibinfo{year}{1997}).

\bibitem[{\citenamefont{Helfrich}(1973)}]{Helfrich1973}
\bibinfo{author}{\bibfnamefont{W.}~\bibnamefont{Helfrich}},
  \bibinfo{journal}{Z. Naturforsch.} \textbf{\bibinfo{volume}{28c}},
  \bibinfo{pages}{693} (\bibinfo{year}{1973}).

\bibitem[{\citenamefont{Seifert}(1995)}]{Seifert1995}
\bibinfo{author}{\bibfnamefont{U.}~\bibnamefont{Seifert}}, \bibinfo{journal}{Z.
  Phys. B} \textbf{\bibinfo{volume}{97}}, \bibinfo{pages}{299}
  (\bibinfo{year}{1995}).

\bibitem[{\citenamefont{Manneville et~al.}(2001)\citenamefont{Manneville,
  Bassereau, Ramaswamy, and Prost}}]{Manneville2001}
\bibinfo{author}{\bibfnamefont{J.-B.} \bibnamefont{Manneville}},
  \bibinfo{author}{\bibfnamefont{P.}~\bibnamefont{Bassereau}},
  \bibinfo{author}{\bibfnamefont{S.}~\bibnamefont{Ramaswamy}},
  \bibnamefont{and} \bibinfo{author}{\bibfnamefont{J.}~\bibnamefont{Prost}},
  \bibinfo{journal}{Phys. Rev. E} \textbf{\bibinfo{volume}{64}},
  \bibinfo{pages}{021908} (\bibinfo{year}{2001}).

\bibitem[{\citenamefont{Lomholt}(2006{\natexlab{a}})}]{Mdipole}
\bibinfo{author}{\bibfnamefont{M.~A.} \bibnamefont{Lomholt}},
  \bibinfo{journal}{Phys. Rev. E} \textbf{\bibinfo{volume}{73}},
  \bibinfo{pages}{061914} (\bibinfo{year}{2006}{\natexlab{a}}).

\bibitem[{\citenamefont{Gov}(2004)}]{Gov2004a}
\bibinfo{author}{\bibfnamefont{N.}~\bibnamefont{Gov}}, \bibinfo{journal}{Phys.
  Rev. Lett.} \textbf{\bibinfo{volume}{93}}, \bibinfo{pages}{268104}
  (\bibinfo{year}{2004}).

\bibitem[{\citenamefont{Sankararaman et~al.}(2002)\citenamefont{Sankararaman,
  Menon, and Kumar}}]{Sankararaman2002}
\bibinfo{author}{\bibfnamefont{S.}~\bibnamefont{Sankararaman}},
  \bibinfo{author}{\bibfnamefont{G.}~\bibnamefont{Menon}}, \bibnamefont{and}
  \bibinfo{author}{\bibfnamefont{P.}~\bibnamefont{Kumar}},
  \bibinfo{journal}{Phys. Rev. E} \textbf{\bibinfo{volume}{66}}
  (\bibinfo{year}{2002}), ISSN \bibinfo{issn}{{1539-3755}}.

\bibitem[{\citenamefont{Lacoste and Lau}(2005)}]{Lacoste2005}
\bibinfo{author}{\bibfnamefont{D.}~\bibnamefont{Lacoste}} \bibnamefont{and}
  \bibinfo{author}{\bibfnamefont{A.~W.~C.} \bibnamefont{Lau}},
  \bibinfo{journal}{Europhys. Lett.} \textbf{\bibinfo{volume}{70}},
  \bibinfo{pages}{418–424} (\bibinfo{year}{2005}).

\bibitem[{\citenamefont{Dai et~al.}(1997)\citenamefont{Dai, Ting-Beall, and
  Sheetz}}]{Dai1997}
\bibinfo{author}{\bibfnamefont{J.}~\bibnamefont{Dai}},
  \bibinfo{author}{\bibfnamefont{H.~P.} \bibnamefont{Ting-Beall}},
  \bibnamefont{and} \bibinfo{author}{\bibfnamefont{M.~P.}
  \bibnamefont{Sheetz}}, \bibinfo{journal}{J. Gen. Physiol.}
  \textbf{\bibinfo{volume}{110}}, \bibinfo{pages}{1} (\bibinfo{year}{1997}).

\bibitem[{\citenamefont{Alberts et~al.}(2002)\citenamefont{Alberts, Johnson,
  Lewis, Raff, Roberts, and Walter}}]{Alberts}
\bibinfo{author}{\bibfnamefont{B.}~\bibnamefont{Alberts}},
  \bibinfo{author}{\bibfnamefont{A.}~\bibnamefont{Johnson}},
  \bibinfo{author}{\bibfnamefont{J.}~\bibnamefont{Lewis}},
  \bibinfo{author}{\bibfnamefont{M.}~\bibnamefont{Raff}},
  \bibinfo{author}{\bibfnamefont{K.}~\bibnamefont{Roberts}}, \bibnamefont{and}
  \bibinfo{author}{\bibfnamefont{P.}~\bibnamefont{Walter}},
  \emph{\bibinfo{title}{Molecular biology of the cell}}
  (\bibinfo{publisher}{Garlend Science}, \bibinfo{year}{2002}).

\bibitem[{\citenamefont{Yoshimura and Sokabe}(2010)}]{Yoshimura2010}
\bibinfo{author}{\bibfnamefont{K.}~\bibnamefont{Yoshimura}} \bibnamefont{and}
  \bibinfo{author}{\bibfnamefont{M.}~\bibnamefont{Sokabe}},
  \bibinfo{journal}{J. R. Soc. Interface} \textbf{\bibinfo{volume}{7}},
  \bibinfo{pages}{S307} (\bibinfo{year}{2010}).

\bibitem[{\citenamefont{Charras et~al.}(2004)\citenamefont{Charras, Williams,
  Sims, and Horton}}]{Charras2004}
\bibinfo{author}{\bibfnamefont{G.}~\bibnamefont{Charras}},
  \bibinfo{author}{\bibfnamefont{B.}~\bibnamefont{Williams}},
  \bibinfo{author}{\bibfnamefont{S.}~\bibnamefont{Sims}}, \bibnamefont{and}
  \bibinfo{author}{\bibfnamefont{M.}~\bibnamefont{Horton}},
  \bibinfo{journal}{Biophys. J.} \textbf{\bibinfo{volume}{87}},
  \bibinfo{pages}{2870} (\bibinfo{year}{2004}), ISSN
  \bibinfo{issn}{{0006-3495}}.

\bibitem[{\citenamefont{Batchelder et~al.}(2011)\citenamefont{Batchelder,
  Hollopeter, Campillo, Mezanges, Jorgensen, Nassoy, Sens, and
  Plastino}}]{Batchelder2011}
\bibinfo{author}{\bibfnamefont{E.~L.} \bibnamefont{Batchelder}},
  \bibinfo{author}{\bibfnamefont{G.}~\bibnamefont{Hollopeter}},
  \bibinfo{author}{\bibfnamefont{C.}~\bibnamefont{Campillo}},
  \bibinfo{author}{\bibfnamefont{X.}~\bibnamefont{Mezanges}},
  \bibinfo{author}{\bibfnamefont{E.~M.} \bibnamefont{Jorgensen}},
  \bibinfo{author}{\bibfnamefont{P.}~\bibnamefont{Nassoy}},
  \bibinfo{author}{\bibfnamefont{P.}~\bibnamefont{Sens}}, \bibnamefont{and}
  \bibinfo{author}{\bibfnamefont{J.}~\bibnamefont{Plastino}},
  \bibinfo{journal}{PNAS} \textbf{\bibinfo{volume}{108}},
  \bibinfo{pages}{11429} (\bibinfo{year}{2011}), ISSN
  \bibinfo{issn}{{0027-8424}}.

\bibitem[{\citenamefont{Brannigan et~al.}(2006)\citenamefont{Brannigan, Lin,
  and Brown}}]{Brannigan2006}
\bibinfo{author}{\bibfnamefont{G.}~\bibnamefont{Brannigan}},
  \bibinfo{author}{\bibfnamefont{L.}~\bibnamefont{Lin}}, \bibnamefont{and}
  \bibinfo{author}{\bibfnamefont{F.}~\bibnamefont{Brown}},
  \bibinfo{journal}{Eur. Bio. J. with Bio. Lett.}
  \textbf{\bibinfo{volume}{35}}, \bibinfo{pages}{104} (\bibinfo{year}{2006}),
  ISSN \bibinfo{issn}{{0175-7571}}.

\bibitem[{\citenamefont{Prost and Bruinsma}(1996)}]{Prost1996}
\bibinfo{author}{\bibfnamefont{J.}~\bibnamefont{Prost}} \bibnamefont{and}
  \bibinfo{author}{\bibfnamefont{R.}~\bibnamefont{Bruinsma}},
  \bibinfo{journal}{Europhys. Lett.} \textbf{\bibinfo{volume}{33(4)}},
  \bibinfo{pages}{321} (\bibinfo{year}{1996}).

\bibitem[{\citenamefont{Gov et~al.}(2003)\citenamefont{Gov, Zilman, and
  Safran}}]{Gov2003}
\bibinfo{author}{\bibfnamefont{N.}~\bibnamefont{Gov}},
  \bibinfo{author}{\bibfnamefont{A.~G.} \bibnamefont{Zilman}},
  \bibnamefont{and} \bibinfo{author}{\bibfnamefont{S.}~\bibnamefont{Safran}},
  \bibinfo{journal}{Phys. Rev. Lett.} \textbf{\bibinfo{volume}{90}},
  \bibinfo{pages}{228101} (\bibinfo{year}{2003}).

\bibitem[{\citenamefont{Gov and Safran}(2004)}]{Gov2004b}
\bibinfo{author}{\bibfnamefont{N.}~\bibnamefont{Gov}} \bibnamefont{and}
  \bibinfo{author}{\bibfnamefont{S.~A.} \bibnamefont{Safran}},
  \bibinfo{journal}{Phys. Rev. E} \textbf{\bibinfo{volume}{69}},
  \bibinfo{pages}{011101} (\bibinfo{year}{2004}).

\bibitem[{\citenamefont{Gov and Safran}(2005)}]{Gov2005}
\bibinfo{author}{\bibfnamefont{N.}~\bibnamefont{Gov}} \bibnamefont{and}
  \bibinfo{author}{\bibfnamefont{S.}~\bibnamefont{Safran}},
  \bibinfo{journal}{Biophys. J.} \textbf{\bibinfo{volume}{88}},
  \bibinfo{pages}{1859} (\bibinfo{year}{2005}).

\bibitem[{\citenamefont{Girard et~al.}(2005)\citenamefont{Girard, Prost, and
  Bassereau}}]{Girard2005}
\bibinfo{author}{\bibfnamefont{P.}~\bibnamefont{Girard}},
  \bibinfo{author}{\bibfnamefont{J.}~\bibnamefont{Prost}}, \bibnamefont{and}
  \bibinfo{author}{\bibfnamefont{P.}~\bibnamefont{Bassereau}},
  \bibinfo{journal}{Phys. Rev. Lett.} \textbf{\bibinfo{volume}{94}},
  \bibinfo{pages}{088102} (\bibinfo{year}{2005}).

\bibitem[{\citenamefont{Faris et~al.}(2009)\citenamefont{Faris, Lacoste,
  Pecreaux, Joanny, Prost, and Bassereau}}]{FarisPRL}
\bibinfo{author}{\bibfnamefont{M.~D. E.~A.} \bibnamefont{Faris}},
  \bibinfo{author}{\bibfnamefont{D.}~\bibnamefont{Lacoste}},
  \bibinfo{author}{\bibfnamefont{J.}~\bibnamefont{Pecreaux}},
  \bibinfo{author}{\bibfnamefont{J.-F.} \bibnamefont{Joanny}},
  \bibinfo{author}{\bibfnamefont{J.}~\bibnamefont{Prost}}, \bibnamefont{and}
  \bibinfo{author}{\bibfnamefont{P.}~\bibnamefont{Bassereau}},
  \bibinfo{journal}{Phys. Rev. Lett.} \textbf{\bibinfo{volume}{102}},
  \bibinfo{pages}{038102} (\bibinfo{year}{2009}).

\bibitem[{\citenamefont{Seifert}(1999)}]{Seifert1999}
\bibinfo{author}{\bibfnamefont{U.}~\bibnamefont{Seifert}},
  \bibinfo{journal}{Eur. Phys. J. B} \textbf{\bibinfo{volume}{8}},
  \bibinfo{pages}{405} (\bibinfo{year}{1999}).

\bibitem[{\citenamefont{Lomholt}(2006{\natexlab{b}})}]{MdipoleTh}
\bibinfo{author}{\bibfnamefont{M.~A.} \bibnamefont{Lomholt}},
  \bibinfo{journal}{Phys. Rev. E} \textbf{\bibinfo{volume}{73}},
  \bibinfo{pages}{061913} (\bibinfo{year}{2006}{\natexlab{b}}).

\bibitem[{\citenamefont{Lomholt et~al.}(2011)\citenamefont{Lomholt, Loubet, and
  Ipsen}}]{Michael2011}
\bibinfo{author}{\bibfnamefont{M.~A.} \bibnamefont{Lomholt}},
  \bibinfo{author}{\bibfnamefont{B.}~\bibnamefont{Loubet}}, \bibnamefont{and}
  \bibinfo{author}{\bibfnamefont{J.~H.} \bibnamefont{Ipsen}},
  \bibinfo{journal}{Phys. Rev. E} \textbf{\bibinfo{volume}{83}},
  \bibinfo{pages}{011913} (\bibinfo{year}{2011}).

\bibitem[{\citenamefont{Lin et~al.}(2006)\citenamefont{Lin, Gov, and
  Brown}}]{Lin2006}
\bibinfo{author}{\bibfnamefont{L.~C.-L.} \bibnamefont{Lin}},
  \bibinfo{author}{\bibfnamefont{N.}~\bibnamefont{Gov}}, \bibnamefont{and}
  \bibinfo{author}{\bibfnamefont{F.~L.~H.} \bibnamefont{Brown}},
  \bibinfo{journal}{Chem. Phys. J.} \textbf{\bibinfo{volume}{124}},
  \bibinfo{pages}{074903} (\bibinfo{year}{2006}).

\bibitem[{\citenamefont{Gov}(2007)}]{Gov2007}
\bibinfo{author}{\bibfnamefont{N.}~\bibnamefont{Gov}}, \bibinfo{journal}{Phys.
  Rev. E} \textbf{\bibinfo{volume}{75}}, \bibinfo{pages}{011921}
  (\bibinfo{year}{2007}).

\bibitem[{\citenamefont{Sheetz}(2001)}]{Sheetz2001}
\bibinfo{author}{\bibfnamefont{M.}~\bibnamefont{Sheetz}},
  \bibinfo{journal}{Nat. rev. Mol. cell bio.} \textbf{\bibinfo{volume}{2}},
  \bibinfo{pages}{392} (\bibinfo{year}{2001}), ISSN
  \bibinfo{issn}{{1471-0072}}.

\bibitem[{\citenamefont{Tuvia et~al.}(1998)\citenamefont{Tuvia, Levin, Bitler,
  and Korenstein}}]{Tuvia1998}
\bibinfo{author}{\bibfnamefont{S.}~\bibnamefont{Tuvia}},
  \bibinfo{author}{\bibfnamefont{S.}~\bibnamefont{Levin}},
  \bibinfo{author}{\bibfnamefont{A.}~\bibnamefont{Bitler}}, \bibnamefont{and}
  \bibinfo{author}{\bibfnamefont{R.}~\bibnamefont{Korenstein}},
  \bibinfo{journal}{J. Cell Biol.} \textbf{\bibinfo{volume}{141}},
  \bibinfo{pages}{1551} (\bibinfo{year}{1998}).

\bibitem[{\citenamefont{Vo\"itchovsky et~al.}(2009)\citenamefont{Vo\"itchovsky,
  Contera, and Ryan}}]{Voitchovsky2009}
\bibinfo{author}{\bibfnamefont{K.}~\bibnamefont{Vo\"itchovsky}},
  \bibinfo{author}{\bibfnamefont{S.}~\bibnamefont{Contera}}, \bibnamefont{and}
  \bibinfo{author}{\bibfnamefont{J.~F.} \bibnamefont{Ryan}},
  \bibinfo{journal}{Soft Matter} \textbf{\bibinfo{volume}{5}},
  \bibinfo{pages}{4899} (\bibinfo{year}{2009}).

\bibitem[{\citenamefont{Yang et~al.}(1997)\citenamefont{Yang, van Hoek, and
  Verkman}}]{Yang1997}
\bibinfo{author}{\bibfnamefont{B.}~\bibnamefont{Yang}},
  \bibinfo{author}{\bibfnamefont{A.~N.} \bibnamefont{van Hoek}},
  \bibnamefont{and} \bibinfo{author}{\bibfnamefont{A.~S.}
  \bibnamefont{Verkman}}, \bibinfo{journal}{Biochemistry}
  \textbf{\bibinfo{volume}{36}}, \bibinfo{pages}{7625} (\bibinfo{year}{1997}).

\end{thebibliography}

\end{document}